\documentclass[twocolumn]{aastex63}
\usepackage{lineno}
\usepackage{xcolor}
\usepackage{xspace}
\usepackage{graphicx, natbib, color, bm, amsmath, epsfig, mathrsfs, xfrac, xspace}
\usepackage[inline]{enumitem}
\usepackage{booktabs}
\usepackage{ifthen}
\usepackage{ulem}
\usepackage{acronym}
\newcommand{\myedit}[1]{\textbf{#1}}

\acrodef{AIC}[\myedit{AIC}]{\myedit{``accretion induced collapse''}}
\acrodef{DD}[\myedit{DD}]{\myedit{double degenerate}}
\acrodef{DDT}[\myedit{DDT}]{\myedit{deflagration-to-detonation transition}}
\acrodef{HeD}[\myedit{HeD}]{\myedit{helium detonation}}
\acrodef{IR}[\myedit{IR}]{\myedit{infrared}}
\acrodef{LC}[\myedit{LC}]{\myedit{light curve}}
\acrodef{MHD}[\myedit{MHD}]{\myedit{magnetohydrodynamics}}
\acrodef{MIR}[\myedit{MIR}]{\myedit{mid infrared}}
\acrodef{NIR}[\myedit{NIR}]{\myedit{near infrared}}
\acrodef{NLTE}[\myedit{non-LTE}]{\myedit{non-local thermodynamic equilibrium}}
\acrodef{PDD}[\myedit{PDD}]{\myedit{pulsating delayed detonation}}
\acrodef{QNSE}[\myedit{QNSE}]{\myedit{quasi nuclear statistical equilibrium}}
\acrodef{QSE}[\myedit{QSE}]{\myedit{quasi nuclear statistical equilibrium}}
\acrodef{RT}[\myedit{RT}]{\myedit{Rayleigh-Taylor}}
\acrodef{SD}[\myedit{SD}]{\myedit{single degenerate}}
\acrodef{SN}[\myedit{SN}]{\myedit{supernova}}
\acrodef{SNe}[\myedit{SNe}]{\myedit{supernovae}}
\acrodef{SNIa}[\myedit{SN Ia}]{\ac{SN}~\myedit{Type~Ia}}
\acrodef{SNeIa}[\myedit{SNe Ia}]{\ac{SNe}~\myedit{Type~Ia}}
\acrodef{WD}[\myedit{WD}]{\myedit{white dwarf}}

\def\hydra{\textit{HYDRA}}

\newcommand{\myMCh}{\ensuremath{M_{\rm Ch}}}

\definecolor{orange}{rgb}{1.        ,  0.54,  0}
\definecolor{remove}{rgb}{0.5        ,  0.5,  0.5}

\newcommand{\Msun}{$M_{\odot}$\xspace}

\newcommand{\SiI}{\ion{Si}{1}}
\newcommand{\SiII}{\ion{Si}{2}}

\newcommand{\SI}{\ion{Si}{1}}
\newcommand{\SII}{\ion{Si}{2}}
\newcommand{\CaII}{\ion{Ca}{2}}

\newcommand{\FeI}{\ion{Fe}{1}}
\newcommand{\FeII}{\ion{Fe}{2}}
\newcommand{\FeIII}{\ion{Fe}{3}}

\newcommand{\CoII}{\ion{Co}{2}}
\newcommand{\CoIII}{\ion{Co}{3}}
\newcommand{\NiII}{\ion{Ni}{2}}

\newcommand{\NaI}{\ion{Na}{1}}

\newcommand{\qxp}{SN~2020qxp/ASASSN-20jq\xspace}

\newcommand{\gcm}{g~cm$^{-3}$\xspace}
\newcommand{\kmps}{\ensuremath{\text{km}~\text{s}^{-1}}\xspace}

\newcommand{\kms}{\text{km}$^{-1}$\xspace}
\newcommand{\msol}{\ensuremath{{\text{M}_\odot}}\xspace}

\newcommand{\mum}{\text{µm}} 
\newcommand{\microns}{\text{µm}\xspace}
\newcommand{\feiiprofile}{right-triangle shaped\xspace}

\definecolor{meta}{rgb}{0.371,0.617,0.625} 

\def\nar{New Astronomy Reviews}

\newcommand{\newtextc}[1]{{#1}}

\received{07/26/2021}
\revised{09/03/2021}
\accepted{09/07/2021}
\submitjournal{ApJ}
\begin{document}
\title{Measuring an off-Center Detonation through Infrared Line Profiles: \\ The peculiar Type Ia Supernova SN~2020qxp/ASASSN-20jq}
\begin{abstract}
We present and analyze a near infrared
(NIR) spectrum of the underluminous Type Ia supernova SN~2020qxp/ASASSN-20jq obtained with NIRES at the Keck Observatory 
$191$ days after $B$-band maximum. 
	The spectrum is dominated by a number of broad emission features including the [FeII] at  1.644~\microns  which is  highly asymmetric with a tilted top and
a peak red-shifted by $\approx$ 2,000 km~s$^{-1}$.  In comparison \newtextc{with 2-D non-LTE synthetic spectra computed from 3-D simulations} of off-center
delayed-detonation Chandrasekhar-mass ($M_{\rm{ch}}$) white-dwarf(WD) models, we  find good agreement between the observed lines and the synthetic profiles, and are able to unravel the structure of the progenitor's envelope.
We  find that the size and tilt of the [Fe II] 1.644~\microns profile
(in velocity space) is an effective way to determine the location of an
off-center delayed-detonation transition (DDT) and the viewing angle, and it requires a WD with a high central density of  $\sim 4\times10^{9}$~g~cm$^{-3}$.
We also tentatively identify  a stable Ni feature around 1.9~\microns characterized by a `pot-belly' profile that is slightly offset with  respect to the kinematic center. 
In the case of SN~2020qxp/ASASSN-20jq, we estimate that  the location
of the DDT is $\sim 0.3M_{\rm WD}$  off-center, which gives rise to an asymmetric distribution of the underlying ejecta. We also demonstrate that low-luminosity and high-density WD SN~Ia progenitors
exhibit a very strong overlap of Ca and $^{56}$Ni in physical space. This results in the formation of a prevalent [Ca II] 0.73~\microns emission feature which is sensitive to asymmetry effects. 
Our findings are discussed within the context of alternative scenarios, including off-center C/O detonations in He-triggered sub-$M_{\rm{Ch}}$ WDs and the direct collision of two WDs.  Snapshot programs \newtextc{with} Gemini/Keck/VLT/ELT class instruments and our  spectropolarimetry program are complementary to mid-IR spectra by JWST.
\end{abstract}
\author[0000-0002-4338-6586]{ P.Hoeflich}
\affil{Department of Physics, Florida State University, Tallahassee, Fl 32306, USA}

\author[0000-0002-5221-7557]{C. Ashall}
\affil{Institute for Astronomy, University of Hawaii, 2680 Woodlawn Drive, Honolulu, HI 96822, USA}
\author[0000-0003-3529-3854]{S. Bose}
\affiliation{Department of Physics and Astronomy, Ohio State University, 4004 McPherson Laboratory,  Columbus, Ohio 43210-1173, USA}
\affil{Center for Cosmology and AstroParticle Physics (CCAPP), Ohio State University, 191 W. Woodruff Ave., Columbus, OH 43210, USA}
\author[0000-0001-5393-1608]{E. Baron}
\affiliation{Homer L. Dodge Department of Physics and Astronomy, University of Oklahoma, 440 W. Brooks, Norman, OK 73019-2061, USA}
\author[0000-0002-5571-1833]{M.~D.~Stritzinger}
\affiliation{Department of Physics and Astronomy, Aarhus University, Ny Munkegade 120, DK-8000 Aarhus C, Denmark}
\author[0000-0002-2806-5821]{S. Davis}
\affil{Department of Physics, University of California, 1 Shields Avenue, Davis, CA 95616-5270, USA}
\author[0000-0001-5393-1608]{M. Shahbandeh}
\affiliation{Department of Physics, Florida State University, Tallahassee, Fl 32306, USA}
\author[0000-0003-0805-3343] {G.~S.~Anand}
\affiliation{Institute for Astronomy, University of Hawaii, 2680 Woodlawn Drive, Honolulu, HI 96822, USA}
\author[0000-0003-0805-3343]
{D. Baade}
\affil{European Organization for Astronomical Research in the  Southern Hemisphere (ESO), Karl-Schwarzschild-Str. 2, Garching b. M\"unchen, Germany}
\author[0000-0003-4625-6629]{C.~R.~Burns}
\affil{Observatories of the Carnegie Institution for Science, 813 Santa Barbara St., Pasadena, CA 91101, USA}
\author[0000-0001-6661-2243]{D. C. Collins}
\affiliation{Department of Physics, Florida State University, Tallahassee, Fl 32306, USA}
\author[0000-0002-0805-1980]{T. R. Diamond}
\affil{Private Astronomer: tiaradiamond@gmail.com}
\author[0000-0002-5253-3584]{A. Fisher}
\affiliation{Department of Physics, Florida State University, Tallahassee, Fl 32306, USA}
\author[0000-0002-1296-6887]{L.~Galbany}
\affil{Institute of Space Sciences (ICE, CSIC), Campus UAB, Carrer de Can Magrans, s/n, E-08193 Barcelona, Spain.}
\author[0000-0001-9556-7576] {B. A. Hristov}
\affil{Center for Space Plasma and Aeronomic Research, University of Alabama in Huntsville, Huntsville, Alabama, USA}
\author[0000-0001-9556-7576]{E.Y.~ Hsiao}
\affiliation{Department of Physics, Florida State University, Tallahassee, Fl 32306, USA}
\author[0000-0001-9556-7576]{M.M.~ Phillips}
\affil{Carnegie Observatories, Las Campanas Observatory, Casilla 601, La Serena, Chile}
\author[0000-0002-2806-5821]{B.~ Shappee}
\affil{Institute for Astronomy, University of Hawaii, 2680 Woodlawn Drive, Honolulu, HI 96822, USA}
\author[0000-0001-9556-7576]{N.B.~ Suntzeff}
\affil{P. and C. Woods Mitchell Institute for Fundamental
 Physics and Astronomy, Department of Physics and Astronomy, Texas
 A\&M University, College Station, TX 77843, USA}
 \author[0000-0002-2806-5821]{M.~ Tucker}
\affiliation{Institute for Astronomy, University of Hawaii, 2680 Woodlawn Drive, Honolulu, HI 96822, USA}


\def\myMch{\ensuremath{M_{\rm{Ch}}}}

\section{Introduction}
\label{sect:intro}

   Type~Ia supernovae (SNe~Ia)  are  thermonuclear disruptions   of carbon-oxygen  white dwarf (WD) stars \citep{hf60}.
These cosmic explosions are significant producers of Fe-group elements in the Universe, and when used as  cosmological distance indicators \citep{Pskovskii1977,p93}, they  enable us to map out the expansion history of the Universe  to red-shifts of $z\gtrsim 2$. 

Over the past two decades detailed studies of hundreds of SNe~Ia have
led to the realization of significant  diversity within the population
in terms of both luminosity (a range of a factor of $\sim 10$)  and spectral properties.  
For example, there are high- and low-velocity objects \citep{2005ApJ...623.1011B}, both of which being characterized by a range of  gradients, while spectral line diagnostics has led the identification of various SN~Ia sub-types  \citep{2005PASP..117..545B,2009PASP..121..238B,2013Sci...340..170W,2013ApJ...773...53F}.
The source of these spectral differences may be linked to 
variations in progenitor systems,  explosion scenarios
\citep{hk96,quimby06,shen10,Polin19}, and/or due to viewing angle effects \citep{howell01,2003ApJ...591.1110W,Hoeflich2006,Motohara06,maedanature10,Shen2018}.

Potential  SN~Ia progenitor systems include: (i) a single degenerate
(SD) system consisting of a single WD with a non-degenerate donor
companion  which may be, with increasing orbital separation, a helium (He), main sequence, or red giant star
\citep{iben84,Webbink84,han06a,Stefano11}; (ii) a double degenerate
(DD) system consisting of two WDs in close orbit with velocities $\gtrsim 1500~$ \kmps that merge via  the potential energy loss by
gravitational radiation \citep{iben84,Webbink84} or 
(iii) a triple system with two colliding WDs resulting in a peculiar motion of the center of mass
\citep{lidov62,Rosswog2009,Thompson2011,Pejcha2013,Kushnir2013,Dong2015}.


In addition to the progenitor system, the  explosion physics of  SNe~Ia is highly debated with three leading scenarios. The first is known as the delayed-detonation scenario \citep{khok89}.
A WD accretes material from a companion in either a DD system on long time scales, so-called secular mergers, or in a SD system
\citep{WI73,Piersanti2004}.
The explosion is triggered by compressional heating close to the center of the WD as it approaches $M_{\rm{Ch}}$.
The flame starts as a deflagration \citep{Nomoto84}, followed by a  deflagration to detonation transition, DDT.
 The DDT is likely due the mixing of burned and unburned matter following the so-called Zeldovich mechanism \citep{k05,niemeyer96,Hristov2018,2019Sci...366.7365P,2021MNRAS.501L..23B}.
 Asymmetries in the abundance distribution are to be expected on small scales due to  Rayleigh-Taylor (RT) instabilities and, on large scales, in the case of an off-center DDT.


 A second leading explosion  scenario considers
  a surface He detonation (HeD) that triggers a central detonation of a sub-\myMCh {WD} with a C/O core \citep{wwt80,n82,livne1990,Woosley94,hk96,Kromer2010,Sim10,WK2011,Shen2015,Tanikawa2018,Glasner2018,2019arXiv190310960T}. In reality, though, the C/O detonation may well be triggered off-center \citep{livne05}. Basic
characteristics are low-density burning with little production of electron-capture (EC) elements, and a rather spherical distribution of iron-group elements.


In the third possible scenario, two {WD}s merge or collide, possibly head-on in a triple system \citep{Webbink84,iben84,benz90,rasio94,hk96,segretain97,yoon2007,WMC09,WCMH09,loren09,Pakmor10,isern11,pakmor12,Rosswog2009,Thompson2011,Pejcha2013,Kushnir2013,Dong2015,Garcia-BerroHB}.
This process occurs on a dynamical timescale, much faster than the
slow accretion timescales in secular mergers.
 In simulations of this process, the ejecta show large-scale density and abundance asymmetries.


 We present a  late-phase, medium resolution near-infrared (NIR) 
spectrum of the peculiar Type~Ia \qxp .
Our high-quality spectrum exhibits unique line profiles that enable us to test predictions of leading  progenitor/explosion scenarios. 
 Particular attention is paid to  the 1.644~\microns ~ feature which is formed predominantly by a single [\FeII] line transition, rather than from multiple   [\FeII],[\FeIII] and [\CoII]/[\CoIII] blends that produce the majority of other prevalent  optical/NIR features in late-phase SN~Ia spectra.
 
 Here, we use this nebular phase spectrum and the [\FeII]  profile at 1.644 \microns of an underluminous SN~Ia to develop new methods to provide insight in the explosion physics, and  density (and
mass) of the progenitor, and the properties of the progenitor
system. Special emphasis will be put on the 3D imprint of the ejecta on spectra and their variations with the direction observed. We will use the observed NIR spectrum for verification of the simulations and, as a by-product, establish that the optical [\CaII] line strength plus its profile are a powerful tool to constrain the nature of the explosion.
 
 \begin{figure*}[ht]
    \centering
    \includegraphics[width=1.00\textwidth]{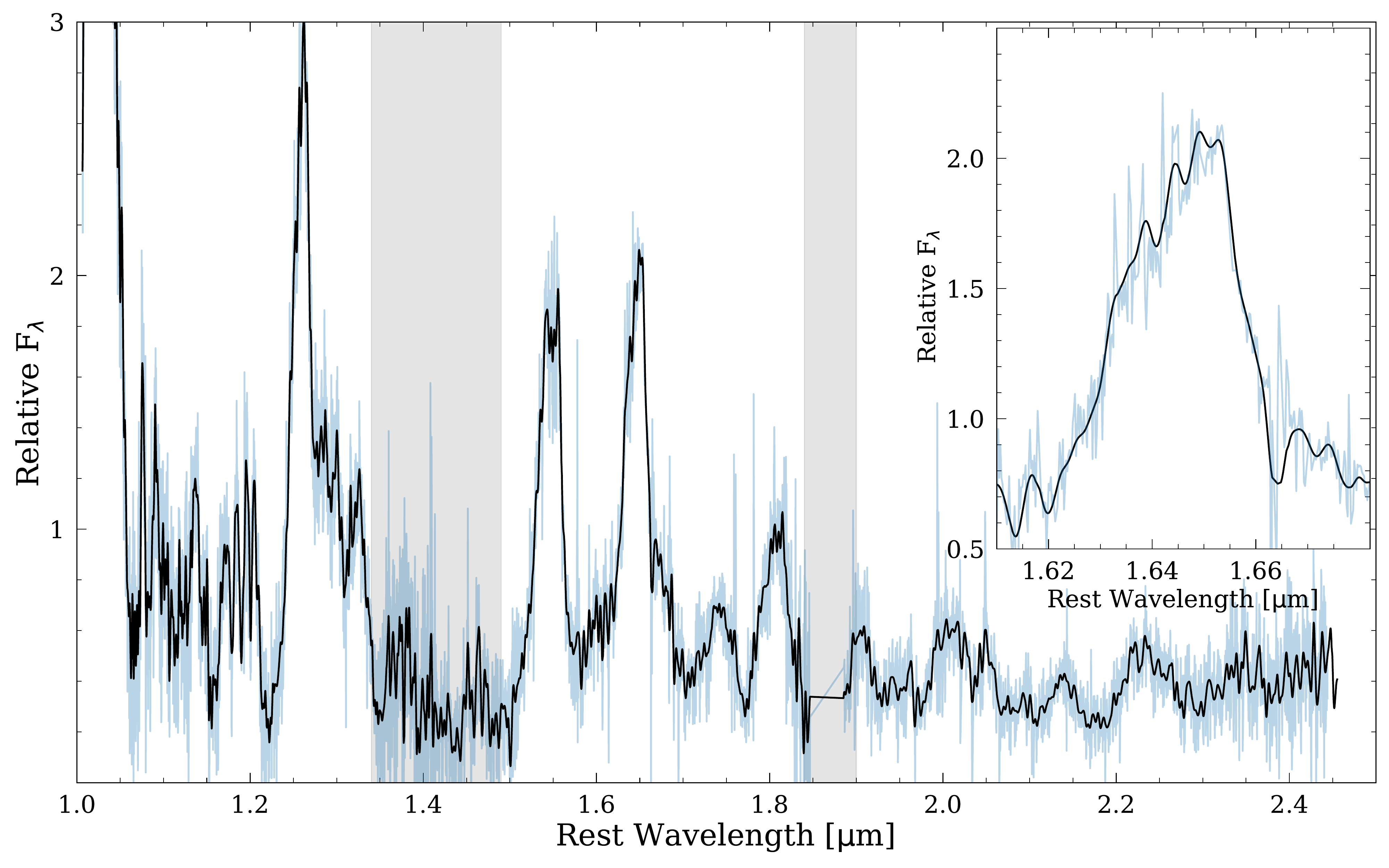} 
    \caption{A  near-IR spectrum of the peculiar type~Ia  \qxp obtained with the Keck telescope (+NIRES) $+$191~d past the epoch of rest-frame $B$-band maximum. The observed 1-D spectrum is plotted in blue, while the black line corresponds to a smoothed version. The inset highlights the unusual  [Fe II]  1.6443~$\micron$ line profile. Shaded
    areas mark regions affected by the atmosphere. Specific line
    transitions are listed in Table~\ref{table.prediction} and
    features are labeled in Figure~\ref{IR_spectrum}.}
\label{fig:data}
\end{figure*}
 
Our paper is organized as follows. 
In \S~\ref{sect:obs} we present the data. In \S~\ref{sect:motiv}, the motivation is given for our approach.
In \S~\ref{sect:models}, our numerical methods are presented to simulate the nebular phase, and the off-center explosion model for \qxp\ is characterized. In \S~\ref{sect:LCP}, we provide estimates for the properties of the early synthetic light curve properties. In \S \ref{sect:Spectral}, we develop the methods to analyze the line profiles and synthetic spectra and apply those to the observation of \qxp. We next present our final discussion in \S~\ref{sect:DC}, followed by our conclusions in \S~\ref{sect:concl}.


\section{Observations of \qxp}
\label{sect:obs}

\qxp\ was discovered by the  All-Sky Automated Survey for Supernovae (ASASSN,  \citealt{Shappee14,Kochanek17}) on 2020 August 8.13~UT in the outskirts of the SBm galaxy NGC~5002.  
The red-shift of the host galaxy is  $z=0.003639$ \citep{2015ApJS..217...27A}\footnote{NASA/IPAC Extragalactic Database (NED)}. There are several direct distance measurements of NGC~5002, all three using the  Tully-Fisher  method. Here we adopt the recent Cosmicflows-4 program  distance modulus $\mu = 31.44\pm0.43$ mag \citep{Kourkchi20}.
The foreground Milky Way reddening along the line-of-sight of \qxp\ is $E(B-V)_{{\rm MW}}$=0.01~mag \citep{Schlafly11}, while the host-galaxy reddening is likely small as there is no evidence for significant \NaI~D in the classification spectrum. 

 \qxp\ was classified initially as a transitional SN~Ia similar to SN~2007on
\citep{Maxi2020}.  Using the classification spectrum,   pseudo-equivalent  width  (pEW) measurements of the  \SiII $\lambda5972$ and \SiII $\lambda$6355 features place \qxp among the cool (CL) SN~Ia subtype on the  Branch diagram.
Furthermore, an initial  analysis of  an unpublished $B$-band light curve (S. Bose et al., in prep) indicates it reached  on  2020 August 23.42 UT (i.e., MJD$59085.43\pm1.52$) an   apparent peak magnitude of   $m_B =14.74 \pm 0.06$~mag.
Correcting for Milky Way reddening and assuming the most recently published Tully-Fischer distance, the peak apparent magnitude corresponds to an absolute magnitude of M$_B = -$16.7$\pm$0.4~mag.
 \qxp is indeed a low-luminosity, CL  SN~Ia.

Our medium-resolution nebular NIR spectrum of \qxp was obtained with the Keck-II telescope equipped with the Near-Infrared Echellette Spectrometer (NIRES) on 2021 March 04.5 UT 
(i.e., MJD=59277.50). This is +191  rest-frame days (d) past the epoch of  $B$-band maximum ($t_B$). The mean resolution of the spectrum is  $R \approx 2700$ and the spectral range extends between $0.96 -  2.46$ $\micron$. Three sets of ABBA exposures were taken, with each individual A/B exposure of 300~s, yielding a  total exposure time on target of 1 hour. The data were  reduced following standard procedures described in the  \texttt{IDL} package Spextools version 5.0.2 for NIRES \citep{spextool}. The extracted 1-D spectrum was   flux calibrated and also corrected for telluric features  with \texttt{Xtellcorr} version 5.0.2 for NIRES making use of observations of the A0V standard star HIP65280 \citep{spextool}. 
As we do not have late-time NIR photometric observations of \qxp\ the absolute \newtextc{flux} scale of our spectrum is uncertain.
However, the line strengths and relative \newtextc{flux} scale is well established due to the reduction method. 

The spectrum of \qxp is plotted in Fig.~\ref{fig:data}. Inspection of the data reveals  broad emission features ubiquitous to normal SNe~Ia at similar epochs.  Most interestingly, the single ion  [\FeII] 1.644~\microns emission feature \citep{h04,Diamond2015}, which is highlighted within the figure's inset, exhibits a  very asymmetric profile and contains several novel properties. 
The 1.644~$\micron$ feature 
  exhibits (going from blue from red) a slanted triangular shape going
  to the peak, followed by an abrupt drop off in the flux redwards of
  the peak. The shape of the profile is similar to that observed in
  the Type~Iax SN~2012Z,  however, the shape is somewhat different and
  rather than describing it as a pot-bellied profile
  \citep{Stritzinger2015}, we refer to it as \feiiprofile,
  where the hypotenuse slopes redward. The peak of the feature is
  located at 1.6545 $\mu$m and is
  therefore   red-shifted  from the rest wavelength of
  1.644~\microns. The top of the feature  is  tilted  over a
  wavelength range corresponding to about 6,000 \kmps in Doppler
  shift. That is, the small leg of the right triangle has a wavelength
  extent corresponding to about 6000~\kmps. Finally, the flux redwards of
  the peak of the feature plunges rapidly to the pseudo-continuum. In
  the following, we will refer to the shape of the [\FeII]
  1.644~\microns feature as a ``\feiiprofile'' profile. Similar profiles
  have previously been produced in the 1.644 [\FeII]~\microns feature
  in our published models
  \citep{h04,penney14,Diamond2015}, but not with the strong
  right-triangle shape seen in \qxp. 

  \section{Motivation}
\label{sect:motiv}
  
Comparisons of well observed SNe~Ia within the same host galaxy and
brightness-decline rates firmly established the diversity in
absolute luminosity and spectra for both normal-bright and
transitional supernovae \citep{Gall2018,BurnsH02018,Burns2020}.
Transitional and sub-luminous SNe~Ia are a key to determining which  scenario(s) truly exist \citep{Gall2018,Ashall18,Galbany2019}.
However, the interpretation of the differences and specific models are
not agreed 
upon. In the case of the transitional twins SNe~2007on and  SNe~2011iv, the
difference of the peak-to-tail ratios of the light curves, colors and photospheric spectra have been
attributed to different central densities of the WD in a $M_{\rm{Ch}}$ mass
explosion \citep{Gall2018}. Alternatively,  based on multi-component
features  in 
the optical spectra, it has also been suggested that SN~2007on, as
opposed to SN~2011iv, resulted from the direct collision of 
two low mass WDs where both WDs were incinerated
\citep{2018MNRAS.476.2905M}. In the case of the sub-luminous SN~2016hnk, 
based on optical light curves and spectra, this object was
attributed to the explosion of a high-density $M_{\rm{Ch}}$ mass WD
because a very narrow (500 km~s$^{-1}$)   
[Ca II] doublet at $0.7293,7326$~\microns  was  observed on top of a wide [Ca II] feature
\citep{Galbany2019}. An alternative interpretation was  a very low
mass, $0.85~ M_\odot$, sub-$M_{\rm{Ch}}$ model that also produced the strong and
wide component of [\CaII],  but  questions the existence of the narrow
[\CaII] component \citep{2020ApJ...896..165J}.  Below, we  discuss a
mechanism to produce even stronger [\CaII] in underluminous  objects
like \qxp\ within a $M_{\rm{Ch}}$ scenario. We show why complex spectra
during the nebular phase depend sensitively on details of the atomic
models, namely non-LTE vs. LTE simulations, and how line profiles add
robust diagnostics.  

We will use the delayed-detonation scenario of a $M_{\rm{Ch}}$ mass WD
because it has predicted many features prior to near- and mid-IR
observations \citep{Wheeler1998,HGFS99by02,Telesco2015}. Moreover, a
significant amount of EC elements
\citep[e.g.][]{brachwitz00} is commonly  detected in X-ray observations of
supernova remnants 
\citep{badenes03,badenes06,thielmann2018,Ohshiro21}, and the solar
isotopic $^{48}$Ca/Ca ratio requires burning under conditions found in
high-density  $M_{Ch} $ WDs \citep{brachwitz00,thielmann2018}. 

Line profiles in the optically-thin nebular phase reveal the asymmetry and burning conditions, and the
imprint of the explosion. Commonly, for spectral features in
the optical and NIR, multi-component  Gaussian fits are employed for
efficiency \citep{Graham2017},  at the cost of a large number of free parameters for each spectral feature. Though guided by atomic
line lists, fits are mostly unrestricted by the underlying physics. \newtextc{The results, nevertheless, are important as they allow identifying elements and main ionization states as a function of time, and this method provides an estimate of the expansion velocities.} Alternatively, 
highly simplified one-zone models, i.e. constant abundances, ionization and temperature are applied without radiation transport \citep{Flors:2020} which severely hampers the link to the explosion physics \newtextc{ because the structure is unlike that of any SNe explosion. However, it shows the importance and validity of atomic data and that, likely, the ionization balance does not change significantly over the line forming region. Semi-analytical methods are complementary to the progress in complex theoretical models. The latter make use of information from both the physical laws and conservation of, e.g., mass and energy, and observations.  Complex models provide a tighter link to the progenitor and explosion physics. Moreover, uncertainties in the underlying physics, e.g., the ionization or atomic data, show up in shortcomings of spectral fits (see \S 4.1 and 6). Detailed methods include } the abundance tomography \citep{Mazzali14}, or non-LTE models using a direct approach \newtextc{
\citep{KF92,h95,JF15,Fransson2015,Diamond2015,Bot2017,Blondin2018,Shingles2020, Wilk2020}}. Therefore, we employ detailed non-LTE model simulations even in multi-dimensional models.   

\begin{table}
\caption{Names and properties of our underluminous delayed-detonation models. We give the central density
$\rho_{\rm c}$ in $10^9 $g~cm$^{-3}$,
the location of the DDT ($M_{\rm WD}$ in $M_{\rm Ch}$),
the peak brightness $B_{max}$ in magnitudes using the Johnson filter systems \citep{bessell90}, and the luminosity to decline ratio 
$\Delta m_{15,s}$ in B and V \newtextc{in magnitudes per day}. All models have the  same main sequence mass,
metallicity and transition density $\rho_{\rm tr}$. \newtextc{ For the spherical models, the DDT is triggered on a sphere with $\rho_{tr}=1.6 \times 10^7$ g/cm$^3$ rather than a point. $M_{DDT}$  is given in parenthesis}.} 
\label{tab:runmatrix}
\begin{center}
\begin{tabular}{|r|rrrr|}
\hline
Model Name&$\rho_{\text{c}}$&$M_{\text{DDT}}$ &$B_{max}$&$\Delta
m_{15,s}(B/V)$\\  
\hline
5p02822d20.16 & 2. & \newtextc{(0.24)} & $-$18.25 & 1.69/1.18 \\
5p02822d40.16 & 4. & \newtextc{(0.25)} & $-$17.92 & 1.42/1.02 \\
Model 00 & 4. & 0.0 &  \nodata & \nodata  \nodata\\
Model 03 & 4. & 0.3 & \nodata  & \nodata  \nodata \\
Model 09 & 4. & 0.9 & \nodata & \nodata  \nodata\\
\hline
\end{tabular}
\end{center}
\label{table0}
\end{table}

\section{Models for the Off-center detonations}
\label{sect:models}

The analysis in this work is based on the DD 
scenario, where the basic parameters come from the model-16-series presented by  \citet{Hoeflich2017}. 
This suite of models has also  been  previously employed in the analysis of the SN remnant S-Andromeda \citep{Fesen07,fesen15}, and for the  transitional type~Ia  SNe~2007on and 2011iv \citep{Gall2018} which exhibit similar brightness characteristics to \qxp. 
The free parameters in the models are the main sequence  mass, $M_{\rm MS}$, metallicity $Z$ of the initial WD, its central density, $\rho_c$, at the time of the explosion, the transition density, $\rho_{{\rm tr}}$, and the location of DDT in mass coordinate $M_{DDT}$. All delayed-detonation models have 
 $M_{MS}= 5$ $M_\odot$, $Z=Z_\odot$, and $\rho_{{\rm tr}}=1.6\times
10^7$~\gcm.\footnote{Note that, within $M_{\rm{Ch}}$ mass explosions, the amount of Si increases with decreasing brightness and, thus, strong Si II and broad-bottom lines are to be expected. The high Si mass is produced on `expense' of $^{56}$Ni. In the 16-series \citep{HGFS99by02}, Si extends to about 20,000 km ~$^{-1}$, but the nuclear quasi-statistical-equilibrium (QSE) region is below 14,000 km~s$^{-1}$.}
We increased $\rho_{\rm c}$ by a
factor of two (i.e., $\rho_{\rm c} =  4 \times 10^9$~\gcm) in order to
reproduce the \feiiprofile profile of the 1.644~\microns feature,
namely to produce the `flat' (linear slope of the central profile) property with the
`tilt' being a result of the off-center DDT as discussed in
\S~\ref{Sect:lineprofiles}, but  keep the other basic parameters
fixed. Our goal is not to produce a specifically tuned model, just to reproduce the overall NIR spectrum of \qxp.


Following the convention adopted by  \citet{dominguez2000,HGFS99by02},
the spherical delayed-detonation models are 
referred to as 5p02822dXX.16  with  XX specifying   $\rho_{\rm c}$ in $10^8$\gcm .

The synthetic spectra are calculated at 210 days after the explosion\footnote{The spherical model has a rise time of 16.7 days.}, which is consistent with $191 \pm 4.3 $ days past the observed $B$ maximum for \qxp (see \S~\ref{sect:obs} \& \S~\ref{sect:LCP}).
 
 In the following, aspherical models are referred to as ``Model YY" with YY denoting the position of the DDT {being the offset in units of  $1/10^{th}$ of the total mass. Note that Model 00 has free parameters identical to the  spherical model 5p02822d40.16 but, for consistency, it was simulated on the 3D grid.} The set of models is shown in Table \ref{tab:runmatrix}.

\noindent
\subsection{Numerical Methods} 
\label{sect:Numerics}
  
Our models are based on full multi-dimensional hydro and non-LTE simulations applied to the nebular phase in SNe. We mention the basic methods, discuss the requirements dictated by the physical conditions, and explain the limitations.
  
The simulations utilized in this work are computed using the
HYDrodynamical RAdiation  code (\hydra ) that consists of
physics-based modules which provide solutions for: the rate equations
that determine the nuclear reactions, the statistical
equations needed to determine the atomic level populations,
the equation-of-state, the matter opacities, 
the hydrodynamic evolution, and the radiation-transport equations (RTE). The RTE is treated by solving the co-moving frame equations in spherical geometry or using Variable Eddington Tensor methods, with a  Monte-Carlo (MC) scheme providing the necessary closure relation to the momentum equations needed to solve the generalized scattering and non-LTE problem.
An MC approach is used for the transport of $\gamma $-rays and positrons and for calculating the emergent spectra \citep{H90,h95,hoeflich2003hydra,Hoeflich:hydro:2009,penney14,Hristov2021}. RTE is needed because most of the SN envelope is still optically thick in the UV \newtextc{(and with significant optical depth in $U$ and $B$)} during the nebular phase, which affects the lower levels via bound-free and allowed bound-bound transitions  \newtextc{and, thus, the ionization balance via the incomplete Rosseland cycle, and the excitation of higher levels.}

In these simulations for the early phase, the hydrodynamical equations
are solved using an explicit Piecewise Parabolic Method (PPM) without
adaptive mesh refinement \citep{CW_PPM84,fryxell,Fryxell91}, followed
by a phase of free expansion of the envelope \citep{Hoeflich17}. 

A nuclear reaction network is used with rates for  weak, strong and
electromagnetic reactions. The nuclear reaction network is  based on the implementation of \citet{T94a,T94b}, but with updated cross-sections  published in \citet{Cy10}.  For this study,  the reduction of the new EC rates by factors up to 5  \citep[e.g.][]{T94a,Langanke04} compared to the old values is important, as it changes the production of EC-elements. This is reflected in the transition from `flat-topped' \citep{h04} to `pot-bellied' profiles \citep{h04,Motohara06,maeda11,Stritzinger2015,Diamond2015,Gall2018}. The revised cross-sections enter directly our estimates of the initial central density required (\S~\ref{sect:Spectral}).

\newtextc{Detailed atomic models are used for the ionization stages I-IV for C, O, Ne, Mg, Si, S, Cl, Ar, Ca, Sc, Ti, V, Cr, Mn, Fe, Co, Ni.\footnote{For zones with particle abundances less than $10^{-5}$, the element is omitted.}
}
The atomic models and line lists are based on the database for bound-bound transitions of \citet{vanHoof2018}\footnote{Version v3.00b3 \url{https://www.pa.uky.edu/~peter/newpage/}},
supplemented by additional forbidden lines from \cite{Diamond2015}.
Note that the atomic levels are based on levels both with and without known transition probabilities because collisions are crucial to avoid
the IR catastrophe \citep{axelphd80,fransson94}, and  because of its apparent absence in observations. This is important to simulate the effective critical density because, in our $M_{Ch} $ mass explosions, the densities at 210 days past the explosion are $\approx 10^7$~particles~cm$^{-3}$ in the region of EC elements.

Though we do not merge fine-structure levels related to the 1.644~\microns ~ feature \citep{penney14,Diamond2015}, the use of superlevels, i.e. the merging of some atomic fine-structure levels, tends to overestimate emission from the excited levels within multiplets of iron-group elements (\S~\ref{sect:overallIR}). 
\newtextc{We assume the radioactive decay by $^{56}$Ni $\rightarrow ^{56}$Co $\rightarrow ^{56}$Fe.} The energy deposition by hard $\gamma $-rays and non-thermal electrons and positrons enters the rate equations via non-thermal ionization balanced by the recombination processes \newtextc{similar to \citet{KF92}, under the assumption that the local nuclear energy input per time is balanced by the local flux \citep{hkm92,h04,penney14,Hristov2021}. For a summary
and some discussion of the  ionization structure, see Appendix \ref{appendix:A}.}

The multi-dimensional non-LTE simulations presented in this paper are limited in resolution.
The overall model spectra are low resolution ($R=150-300$) because memory
requirements for the hydro, radiation transport, and non-LTE atoms
limits the resolution currently possible in \hydra .   3-D models use
(330x70x70) as spatial grid ($r,\Theta,\phi$) and $\approx 6,000$
energy bins for the radiation field. The atomic rate equations are solved 
on ($330\times70$) spatial cells 
using the overall symmetry of the problem discussed in this paper.
For the emergent spectra, we use up to 10,000 frequency counters in the observer's frame in 21 directions.

During the nebular phase, the optical and IR flux is predominately produced by forbidden transitions in the semi-transparent regime and the emissivity is dominated by spontaneous emission providing stability of the solution of the full RTE problem (the solution of the radiative transfer equation, the solution of the rate equations, and conservation of energy) and of the robustness of the spectra. 
\newtextc{In particular, the forbidden unblended [Fe II] at 1.644 \microns line is formed in Fe dominated layers which show rather small variations in the ionization balance. Moreover, being unblended minimizes radiative coupling in an expanding envelope. Combined, this leads to the stability of the resulting line profile as a probe of geometry for features dominated by single transitions. Note that the agreement between the line strenghts of different ions is generally good (see \S 6.2) which might indicate that the predicted ionization and excitation states are close to correct.}
However, atomic data and even some  level energies
\citep{Friesen2017,Diamond2018} are a major uncertainty to reproduce complex profiles of features formed by multiple lines, as discussed in \S~\ref{sect:Spectral}.

\noindent
\subsection{Model Setup}
\label{sect:model_setup}

In constructing off-center delayed-detonation models, we have followed the prescription of
\citet{livne99}, also the approach utilized by \citet{fesen15}.
We limited the deflagration burning to 0.95 of the fuel to allow for the detonation front traveling through the core to bring our results
in close agreement with those of \citet{Gamezo05} and \citet{fesen15}.  In this prescription, the initial deflagration phase is
modeled assuming spherical symmetry.  The deflagration begins at the center and propagates outward in a subsonic deflagration front.  The energy deposited
by deflagration causes the entire WD to expand with a
velocity of  $\sim 2,000-3,000$~\kmps \citep{Hoeflich2017book}.  When the density at
the leading edge of the deflagration wave has fallen to the transition density $\rho_{tr}$, a detonation is ignited by hand at a single point (the north pole) of the deflagration front, imposing rotational symmetry. 

\begin{figure*}[ht]
\begin{minipage}[c]{0.95\textwidth}
  \centering
\includegraphics[scale=0.6]{5p02822d40_16.png}
\end{minipage}
\begin{minipage}[c]{0.80\textwidth}
  \centering
\includegraphics[scale=0.7]{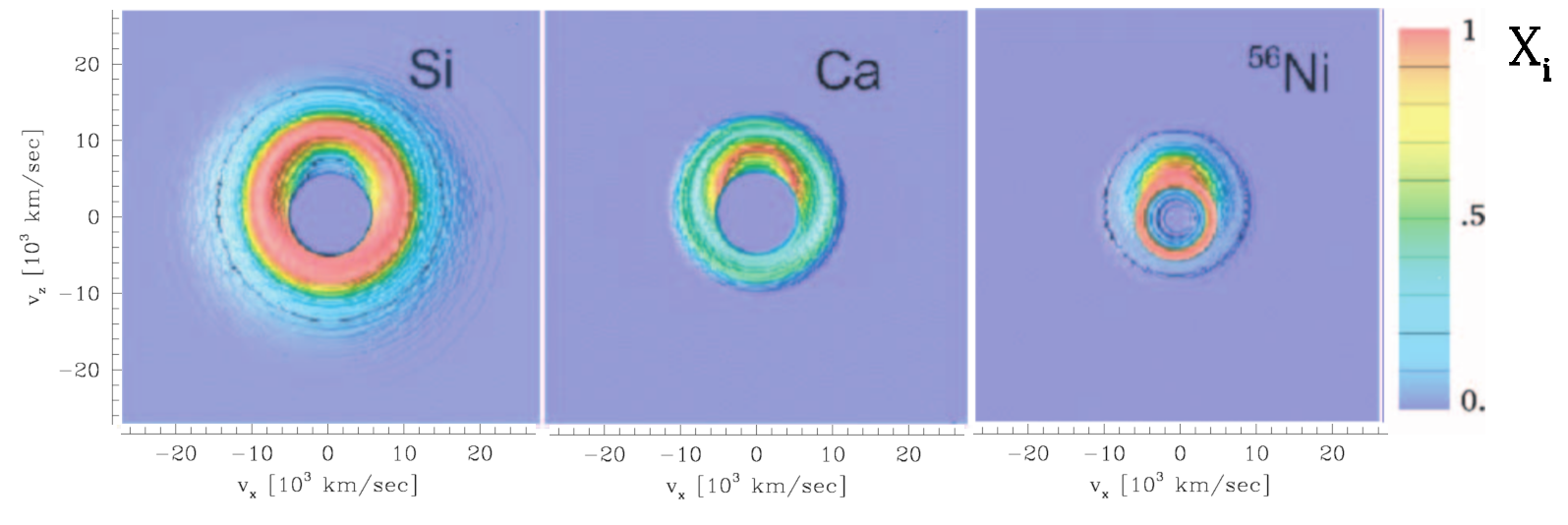}
\end{minipage}
\caption{Distribution of elements in spherical geometry (upper) for our underluminous models with $\rho_c=4 \times 10^9$ g cm$^{-3}$ leading to a region of EC elements of 
$\approx 3,000$ km s$^{-1}$ similar to SN~2016hnk \citep{Galbany19}. 
In the lower panels, we give the distribution of Si, Ca and
$^{56}$Ni for the off-center DDT at $0.3$ $M_{{\rm WD}}$ in velocity coordinates. The abundances are given in mass fractions $X_i$ and are color coded in a domain size of
$\pm 27,500$~\kmps. The inner EC  region is
  spherical as it is produced during the deflagration burning phase
  and shows the same abundances as in the spherical model. Without
  mixing, the size of the EC region shrinks to $\approx 1,500$ and
  $\approx 500$~km s$^{-1}$ for $\rho_c $ of $2 \times 10^9$~\gcm and
  $1 \times 10^9$~\gcm, respectively \citep[see][their
    Fig.~4]{Diamond2015}.}
\label{models}

\end{figure*}

Within the off-center delayed-detonation models, abundance asymmetry is produced
because the DDT occurs on the background of an
already expanding envelope. 
The final outcome of the density structure is close to spherical because the detonation occurs in the density structure of a quasi-static WD.
After the DDT, a weak detonation front propagates through the WD at close
to the speed of sound. Close to the time of the transition, burning occurs under higher density than other layers at the same distance from the center because the detonation reaches those later. 
Subsequently, the hot WD accelerates into free expansion, and the stored thermal energy is converted into kinetic energy. The resulting density structure is close to spherical because the specific kinetic energy hardly depends on whether burning continues to nuclear statistical equilibrium  (NSE)  or quasi-statistical equilibrium (QSE) as also suggested by the full 3D simulation of \citet{Gamezo05}. The final abundance profiles  of Model 03 are shown in Fig.~\ref{models}.

We impose spherical symmetry on the initial deflagration phase for
two reasons:  (i) computational tractability,
3-dimensional simulations are computationally expensive, requiring many $10^5$
CPU hours,
e.g. \citep{Khokhlov01,Roepke03,Gamezo03,Gamezo05,plewa2007,Roepke07,Hristov2021};
(ii) pure hydrodynamical multi-dimensional simulations predict mixed chemical profiles, which are at odds with observations of typical SNe~Ia. 
Observations strongly suggest the existence of a process that partially suppresses the dominant role of Rayleigh-Taylor (RT) instabilities, despite evidence for structures on the scale of RT instabilities.
 These observations include:  post-maximum spectra in normal-bright and sub-luminous SNe~Ia \citep{Wheeler1998,HGFS99by02,Ashall19b};
line profiles in many SNe~Ia  $1-2$ years after the explosion, which indicate stable isotopes near the center after the initial phase of burning
\citep{maeda11,h04,Motohara06,maedanature10,maeda11,stritzinger14b,Diamond2015,Friesen2017,Diamond2018}; direct imaging of S-Andromeda which shows a Ca-free core \citep{Fesen07}.  Recently, high-resolution spectropolarimetry obtained by the VLT  \citep{patat05,Yang2020} shows indications of chemical plumes on the scale of RT instabilities but limited to specific layers in the WD.

The actual origin of this partial suppression of RT instabilities is
not known; however, 
recent 3D simulations of the flame demonstrate the influence of
high magnetic fields on the  deflagration fronts and the  partial
suppression of RT instabilities
\citep{Remming14,Hristov2018}, suggesting a possible physical
mechanism to suppress strong mixing material induced by RT instabilities. 
 The presence of high magnetic fields has been inferred from the lack of positron transport effects in late-time NIR line profiles and light curves which show decline rates that never fall below the $^{56}$Co decay lines \citep{Cappellaro1997, stritzinger02,h04,penney14,Kerzendorf2014,Diamond2015,Graur2017,Shappee2017,Diamond2018,Yang2018,Hristov2021}. 

\section{Light Curve Parameters}
\label{sect:LCP}

The early light curve has been calculated for the 
spherical model 5p02822d40.16 with 
increased central density to $\rho_c = 4\times 10^9$~\gcm. Using the
Johnson filter system \citep{bessell90}, the light curve values for
our high-density version, model 5p02822d40.16, are $M_B=-17.92$~mag,
$B(t_B)-V(t_B)=0.08$~mag with a rise time $t_B=16.9~d$, 
$dm_{15,B}=1.42$, $dm_{15,V}=1.02$~mag respectively for the spherical
model, compared to, with no host reddening correction, $ M_B=
-16.7\pm$0.43~mag for  \qxp (see \S~\ref{sect:obs}). 
 Note that, compared to 5p02822d20.16 with the lower $\rho_{\rm c}$, the high-density model is dimmer, the rise time is shorter and the decline is less steep (see Tab. \ref{table0}) because of the lack of energy deposition close to the center at early times, resulting in shorter diffusion timescales and less energy being stored before maximum light.  Moreover, there is significant uncertainty due to asphericity effects. We did not calculate the time-dependent 3D models during the
photospheric phase. The aspherical Model 03 seen from $-90^o$ will be dimmer
because it is observed from the opposite side of the $^{56}$Ni-bulge
which is hidden during the photospheric phase (see
Fig.~\ref{models}). Stationary scattering models give a factor
$\approx 2$ in luminosity  (see \citet{h95pol} in their Fig. 2, left plots).
We note that the fluxes and properties of the  outer layers are very
sensitive to small variations in the transition density $\rho_{tr}$ in
the regime  of low-luminosity SNe~Ia whereas the inner layers depend
mostly on $\rho_c$ \citep{HGFS99by02,Hoeflich2017}. In light of the
high asymmetry inferred below, fine-tuning of the model parameters has
not been performed, nor is it necessary. The early lightcurve properties from spherical non-LTE simulations
should be regarded as indicative for this class of models. The full 3D
non-LTE lightcurves and spectra during the photospheric phase are currently prohibitively expensive in both CPU and memory requirements.  

\section{Analysis and Interpretation of Line profiles and spectra}
\label{sect:Spectral}
\begin{figure*}[t]
\includegraphics[width=0.99\textwidth]{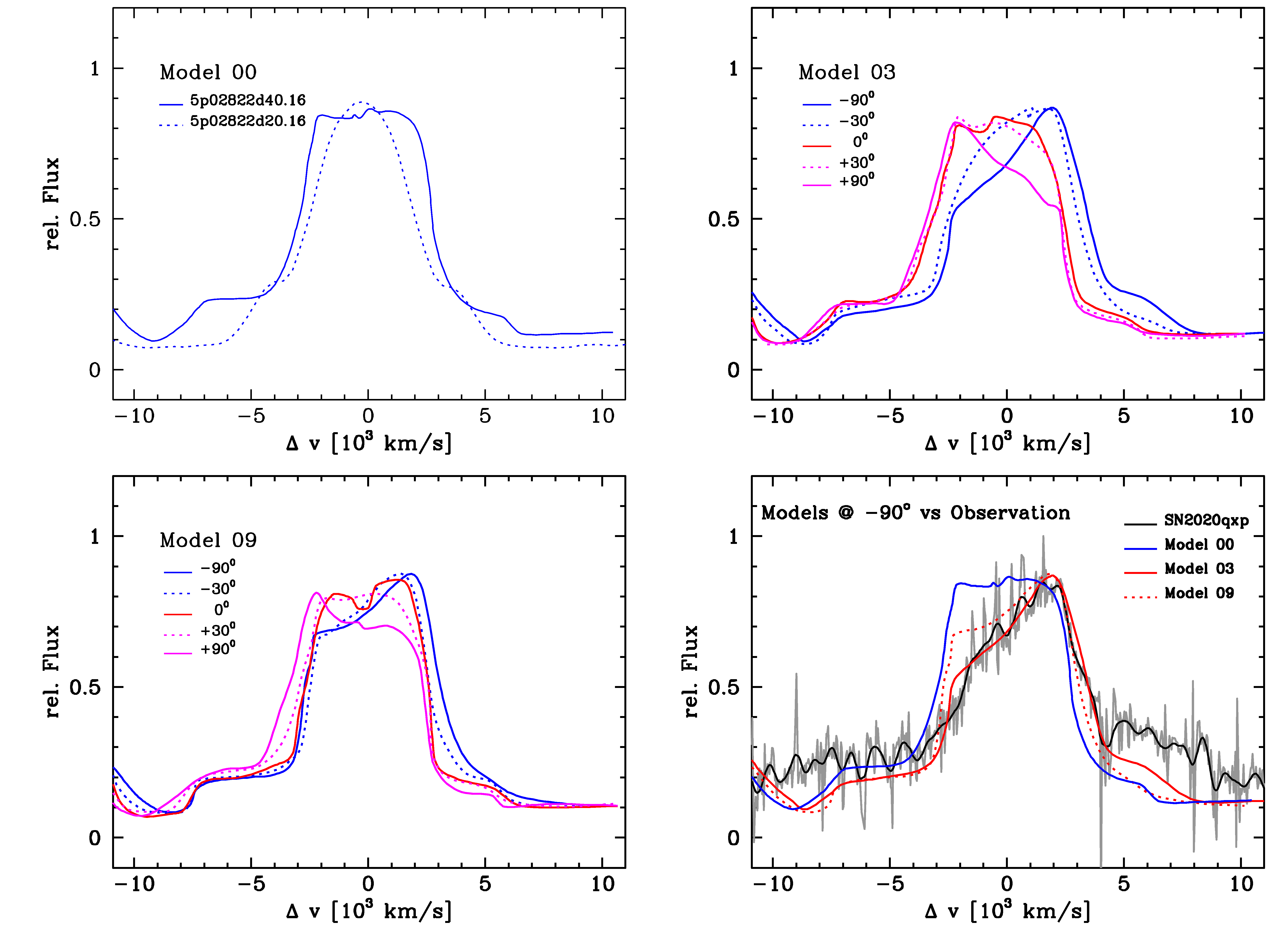}\\
\vskip -0.6cm
\caption{Directional dependence of the synthetic line profiles for off-center delayed-detonation models (upper and lower left), and the comparison with \qxp\ to identify the optimal location of the DDT  (lower right).
The line profile of the minimally blended feature produced by the forbidden [Fe II] at 1.644~\microns at about 210 days after the explosion is shown for our off-center delayed-detonation models seen from various directions, and \qxp. The observation has been blue-shifted by $500 ~\kmps$ to be aligned with the red wing and the center of the flat portion with the
kinematic center of Model 03. The flux peaks at $2,000~ \kmps$. No zero-point adjustments has been applied to the observation, though a 10 \% shift would be compatible with the noise in the data, and improve the fit in the wings beyond $\pm 5,000~\kmps$.
The synthetic spectra are based on our underluminous explosion models with a central density of $\rho_c = 4\times 10^9$~\gcm
with ignition points for the delayed-detonation transitions in mass coordinates at 0., 0.3 and 0.9 $M_{\rm{Ch}}$, Model 00 plus the low-density version with $\rho_c=2\times 10^9$~\gcm (upper left), 03 (upper right) and 09 (lower left).  For a discussion, see \S~\ref{Sect:lineprofiles}. 
 Note that the total width of the synthetic emission component is wider than observed by $\approx 650$ km s$^{-1}$. This can be attributed to the much better resolution of the spectrograph ($\bar{R}=2700$) compared to the effective resolution $R \approx 300 $ of the theoretical model (see \S~\ref{sect:Numerics}).}
\label{IR}
\end{figure*}

\begin{table*}
\begin{center}
    \caption{List of forbidden lines that significantly contribute to the synthetic spectral flux.
 For each transition, the markers correspond to strong ($***$), moderate ($**$), weak ($*$), and 
scarcely
detectable ($~$) on top of the quasi-continuum formed by a large number of lines. \newtextc{ The relative strength S is estimated by the integral over the envelope,   $\int{ A_{ij} \times n_j ~dV }$ with $n_j$ being the particle density of the upper level}. We give the rest wavelength in $\mu$m,
and identify the ion. 
All spectral features are strongly blended with the exception of that due to [\FeII] at $ 1.644$ \microns because the Doppler shift smears out each transition by about $3\%$. \newtextc{Note that transitions are ordered with increasing wavelength from left to right.}}
    \label{table.prediction}
    \begin{tabular}{|rr|rr|rr|rr|rr|}
    \hline
S~~~$\lambda $[\microns] &Ion  &   
S~~~ $\lambda $[\microns]  &Ion  &   
S~~~ $\lambda $[\microns]  &Ion  &   
S~~~  $\lambda $[\microns] &Ion  &     
S~~~  $\lambda $[\microns]&Ion  \\   
\hline
$**$         .9952   & [\FeII]  &   
$ *** $         .9959  & [\NiII]  &   
             .9962   & [\FeIII]  &  
             1.000   & [\FeI]  &    
  $**$         1.001   & [\FeII]  \\   
  $**$         1.004   & [\FeII]  &   
  $**$         1.013  & [\FeII]  &   
  $**$         1.019  & [\CoII]  &   
  $**$         1.021  & [\NiII]  &  
             1.023  & [\FeI]  \\
$ *** $         1.024   & [\CoII]  &   
             1.026  & [\FeI]  &  
 $ *** $         1.028   & [\CoII] &
  $**$         1.029   & [\SII]  &    
  $**$         1.032   & [\SII]  \\    
 $ * $        1.032   & [\SII]  &    
  $**$         1.034  & [\SII]  &    
 $ * $        1.037  & [\SII]  &    
 $ * $        1.043   & [\FeII]  &   
  $**$         1.046   & [\NiII]  \\   
 $ * $        1.057   & [\FeII]  &   
 $ * $        1.064   & [\NiII]  &   
 $ *** $         1.071   & [\NiII]  &   
  $**$         1.082   & [\SI]  &     
  $**$         1.089   & [\FeII]  \\   
 $ * $        1.092   & [\NiII]  &  
  $**$         1.099  & [\SiI]  &   
  $**$         1.104   & [\FeII]  &  
             1.106   & [\CoIII]  & 
  $**$         1.116   & [\FeII]  \\  
  $**$         1.131  & [\SI]  &    
 $ * $        1.135   & [\CoIII]  & 
 $ * $        1.135   & [\FeII]  &  
  $**$         1.136   & [\NiII]  &  
 $ * $        1.146  & [\NiII]  \\  
 $ * $        1.154   & [\SI]  &    
  $**$         1.161   & [\NiII]  &  
              1.187  & [\CoIII]  &
  $**$         1.191  & [\FeII]  &
  $**$         1.205  & [\FeII]  \\
   $*$         1.223  & [\FeII]  &
  $**$         1.232   & [\NiII]  &
  $*$          1.248  & [\FeII]  &
  $**$         1.249  & [\FeII]  &  
  $**$         1.25   & [\FeII]  \\
  $*$          1.252  & [\FeII] &
  $ *** $         1.257   & [\FeII]  &  
 $ *** $         1.271  & [\FeII]  &  
 $ * $        1.273  & [\CoIII]  & 
 $ * $        1.278   & [\NiII]  \\  
 $ *** $         1.279   & [\FeII]  &  
             1.293  & [\NiII]  &  
 $ *** $         1.294   & [\FeII]  &  
  $**$         1.298   & [\FeII]  &  
 $ * $        1.310   & [\CoIII]  \\ 
 $ *** $         1.321  & [\FeII]  &  
             1.321  & [\FeI]  &   
 $ *** $         1.328   & [\FeII]  &  
             1.335   & [\NiII]  &  
 $ * $        1.339   & [\NiII]  \\  
             1.342  & [\FeI]  &    
 $ * $        1.356  & [\FeI]  &    
 $ * $        1.368  & [\FeI]  &    
 $ *** $         1.372   & [\FeII]  &   
 $ * $        1.373   & [\FeI]  \\    
             1.3746 & [\FeI]  &    
 $ * $        1.443   & [\FeI]  &    
  $**$         1.484  & [\FeII]  &   
  $**$         1.491   & [\FeII]  &   
  $**$         1.525   & [\FeII]  \\   
 $ *** $         1.534  & [\FeII]  &   
  $**$         1.547   & [\CoII]  &   
 $ *** $         1.549  & [\CoIII]  &  
             1.588  & [\SiI]  &    
 $ *** $         1.600&[\FeII]  \\   
 $ * $        1.607   & [\SiI]  &    
 $ *** $         1.644   & [\FeII]  &   
  $*$         1.646 & [\SiI]  &    
 $  $        1.646 & [\SiI]  &    
  $**$         1.664   & [\FeII]  \\   
 $ * $        1.677   & [\NiII]  &   
 $ *** $         1.677   & [\FeII]  &   
  $**$         1.712  & [\FeII]  &   
 $ * $        1.725  & [\NiII]  &
 $ * $        1.741   & [\CoIII]  \\  
  $**$         1.745   & [\FeII]  &   
 $ * $        1.764   & [\CoIII]  &  
  $**$         1.765   & [\NiII]  &   
  $**$         1.798  & [\FeII]  &   
  $**$         1.801  & [\FeII]  \\   
 $ * $        1.803  & [\FeII]  &   
  $**$         1.810 & [\FeII]  &   
  $**$         1.812 & [\FeII]  &   
 $ * $        1.814 & [\FeII]  &   
             1.821   & [\CoIII]  \\ 
 $ * $        1.844   & [Mn II]  &   
 $ * $        1.914   & [\FeII]  &   
  $**$         1.939   & [\NiII]  &   
 $ * $        1.958   & [\CoIII]  &  
  $**$         1.967  & [\FeII]  \\   
             1.969   & [\FeIII]  &  
 $ * $        2.002  & [\CoIII]  &  
 $ * $        2.003 & [\FeII]  &   
 $ *** $         2.007   & [\FeII]  &   
             2.022   & [\FeIII]  \\  
 $ *** $         2.047  & [\FeII]  &   
  $**$         2.049   & [\NiII]  &   
 $ * $        2.072  & [\FeII]  &   
 $ * $        2.081   & [\NiII]  &   
 $ * $        2.086   & [\FeII]  \\   
 $ * $        2.098  & [\CoIII]  &  
 $ * $        2.102  & [\NiII]  &   
 $ ** $         2.133   & [\FeII]  &   
             2.146  & [\FeIII]  &  
 $ * $        2.219  & [\FeIII]  \\  
             2.243  & [\FeIII]  &  
 $ *** $         2.244   & [\FeII]  &   
  $**$         2.267  & [\FeII]  &   
 $ * $        2.281   & [\CoIII]  &  
  $**$         2.309  & [\NiII]  \\   
 $ * $        2.335   & [\NiII]  &   
             2.349  & [\FeIII]  &  
 $ * $        2.361   & [\NiII]  &   
  $**$         2.369   & [\NiII]  &   
  $**$         2.371  & [\FeII]  \\   
\hline
\end{tabular}
\end{center}
\label{table2}
\end{table*}

\begin{figure*}[t]
\includegraphics[width=1.05\textwidth]{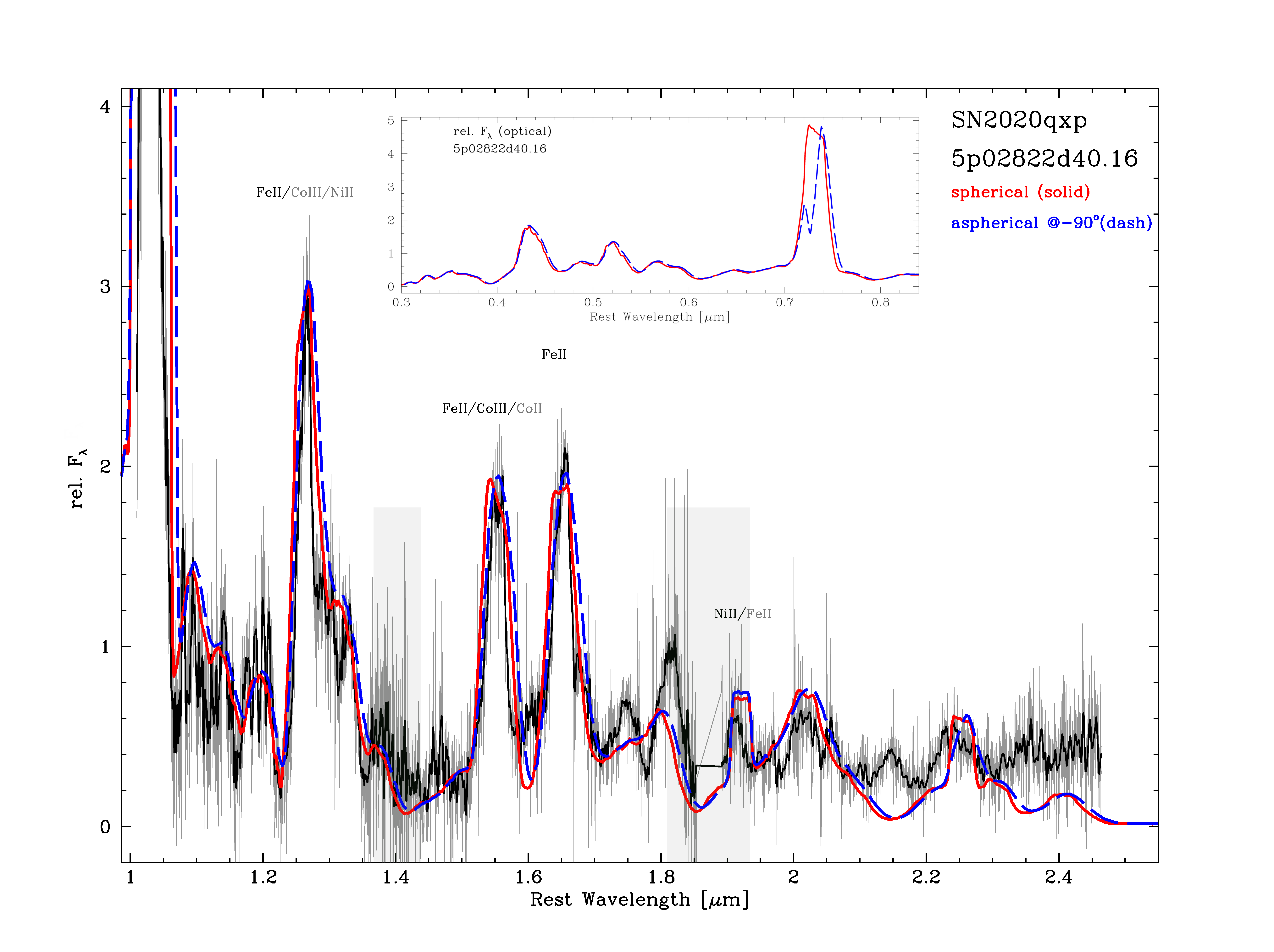}
\vskip -0.5cm
\caption{
Comparison of the synthetic spectrum for the spherical (red) and aspherical model 03 seen from $-90^o$ (blue dashed, $R\approx 250$)  with the NIR Keck spectrum of \qxp from 03/04/2021. 
As an inset, the synthetic optical spectrum is given normalized to the photometric B-band.
The spectrum is dominated by blends of forbidden lines of singly and doubly ionized Fe, Co and Ni, and a weak [Ni II]/[Fe II] blend at about 1.9~microns (see Table \ref{table.prediction}). The three strongest NIR \& the Ni feature are labeled with the main (black) and secondary (gray) contributing ions. Note that the strong
 features at $1.25$ and $1.55~\mu$m have shapes  caused by blends of more than 15 and 7 transitions, respectively, rather than revealing any underlying asymmetries. The [Ni II] feature at $1.9~ \mu$m is flat-topped. The strengths of features are hardly affected by asphericity but Model 03 produces Doppler shifts of the peak fluxes which closely match the observations. For a discussion including the optical [Ca II] feature at 0.73 $\mu$m, see \S~ \ref{sect:overallIR}, and identifications for the strongest optical transitions are given in the appendix. Though observed and theoretical features are consistent in between 1.38-1.44 and 1.81-1.93 $\mu$m (shaded in light gray), those telluric regions are impacted in the observations (see also Fig. \ref{fig:data}).
}

\label{IR_spectrum}
\end{figure*}

We  use the $1.644$~\microns  feature of \qxp\  to find parameters for the central density of the WD, the off-center position of the DDT, and
the inclination angle. Subsequently, we compare the entire NIR
spectrum with these parameters for verification and further analyses.

\begin{figure*}[t]
\includegraphics[width=0.99\textwidth]{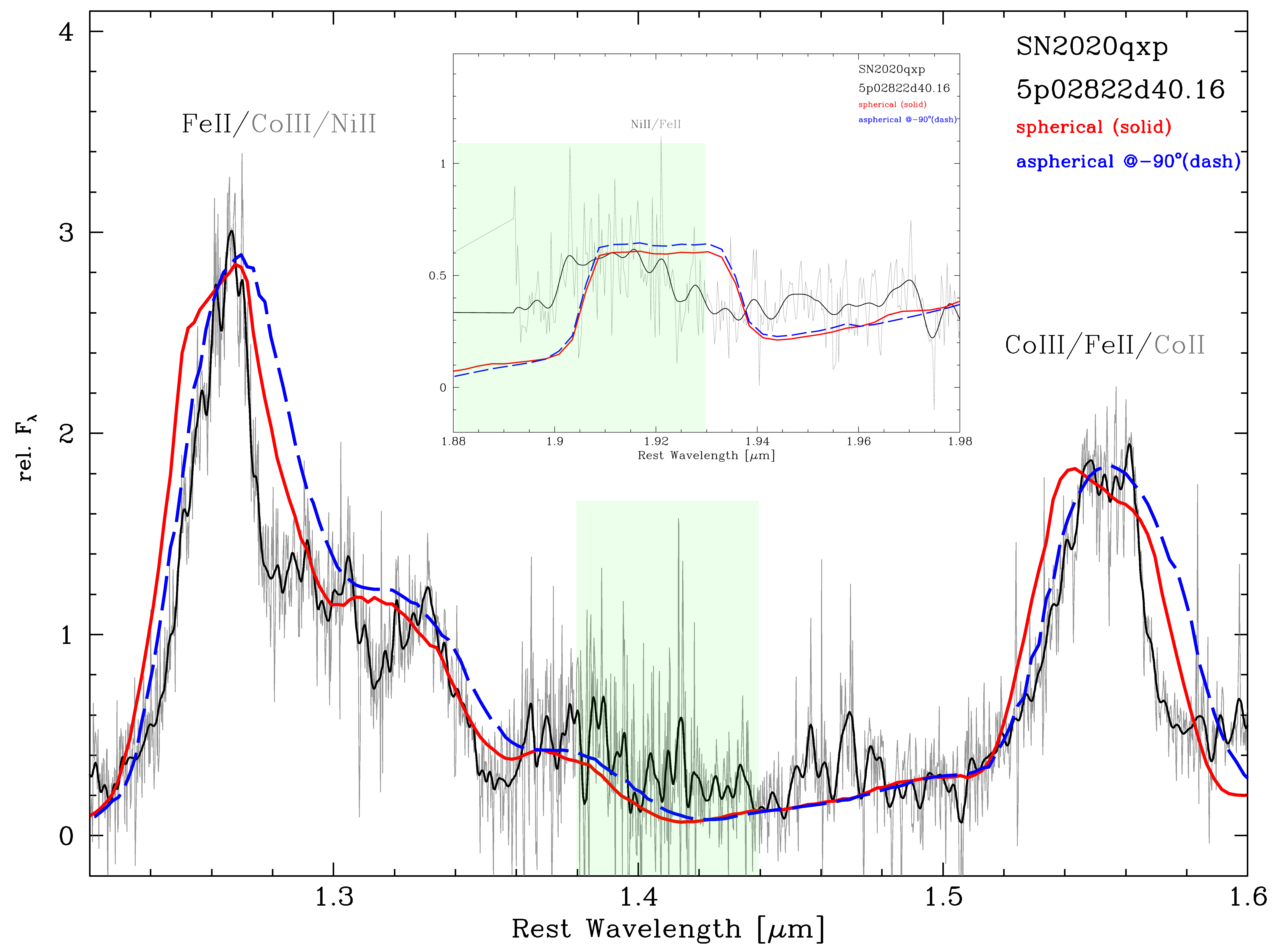}
\vskip -0.1cm
\caption{
  Same as Fig. \ref{IR_spectrum} but with
the blown up spectral regions of the $1.25$~\microns (consisting of many
[\FeII] blends) feature, $1.55$~\microns   (dominated by two [\CoIII]  and [\FeII]
transitions) feature and, as an inset, the $1.9$~\microns feature which is dominated by [\NiII] and weaker [\FeII] transitions (see Table \ref{table.prediction}). The peaks are better reproduced by Model 03. Model 00 tilts the $1.25$ and $1.55$~\microns profiles in opposite direction compared to the observations, and is not a good fit. Neither of these features is dominated by the asymmetry. The $^{58}$Ni feature is flat topped and offset (see \S \ref{Sect:lineprofiles}). 
 Note that the impact of differences in  resolution between observation and synthetic spectrum, and uncertainties related to the atomic models (too strong emission in the red wings from excited fine-structure levels within a multiplet) when describing complex
 features (see \S \ref{sect:Numerics} \& \ref{sect:overallIR}).
 The $1.55$~\microns double-horn profiles produced by [\CoIII] and
 [\FeII] (Table \ref{table.prediction}) are seen in our
 high resolution simulations
 \citep[$R \gtrsim 1,000$][and the observation
   $R=2,700$]{penney14,Diamond2015}, but are  
 merged due to the low $R_{\text{eff}} \sim 500$ in our 3D simulations.}
\label{IR_spectrum2}
\end{figure*}

\subsection{Profile of the 1.644 micron Feature} 
\label{Sect:lineprofiles}
This line profile can be understood by the abundance distribution.
In Fig. \ref{IR}, the profiles of the $1.644$~\microns feature are shown for Models 00, 03 and 09, and compared to the profile of \qxp.
The overall profiles are characterized by an almost unblended central region with blue and red shoulders due to the components of weak [\FeII] blends at wavelength shifts corresponding to $\approx -6,000$ and $\approx +5,000 $~\kmps \citep{h04,Diamond2015}. 

We first discuss the model with a central DDT that produces a spherical chemical distribution (Fig. \ref{IR},  upper left). As can be seen, the synthetic profile is `flat-topped' similar to our old model with $\rho_c = 2 \times 10^9$~\gcm  \citep{h04}. As discussed in \S \ref{sect:Numerics}, the reduction of the EC rates in 2006 combined with lower $\rho_c$ $\approx 0.5 - 2 \times 10^9$ \gcm, typically found in SNe~Ia, caused peaked profiles \citep{Hoeflich2006,penney14,Diamond2015,Diamond2018}.
  To produce a flat top with low EC rates, we now require a higher $\rho_c$ by a factor of $\ga 2$. The extent of the flat top can be used to estimate the size of the $^{56}$Ni-free region in velocity space and requires  $\rho_c = 4\times 10^9 $~\gcm.

 The 1.644 $\mu $m profile of the spherical model does not show a fully symmetric emission component, namely the wings at about $\pm 5,000$~\kmps  are due to well known  blends of weak [\FeII] and [\CoIII]  transitions \citep{h04,Diamond2015}. For our model, [\SiI] contributes to the underlying continuum flux between $\approx -8000$ to $9000$~\kmps basically adding to the flux a broad quasi-continuum contribution.
 
Another characteristic is the lack of emission  at the `red' edge of the flat plateau. It is produced in the models because the envelope is not fully optically thin. At day 210 past the explosion, the photosphere is at about $1,000$ to $1,500$ \kmps in the IR, formed by a quasi-continuum of overlapping allowed lines, Thomson scattering, and free-free emission, effectively blocking the light from parts of the redshifted ejecta. 
At this epoch, our model is making the transition from the
photospheric to the nebular phase.
 
 Similarly, the off-center Models 03 and 09 can be understood based on the asymmetric distribution of $^{56}$Ni
 (Fig. \ref{models}, lower right). Asymmetry introduces a tilt of the flat top  (Fig.~\ref{IR}, upper and lower left).
  Obviously, the profile depends on the direction to the observer. Seen from the equator, the profiles are rather similar to Model 00. Seen from the north and south pole, the $^{56}$Ni produced during the detonation adds a blue and red-shifted component, respectively. The length of the plateau in  velocity space remains almost the same, but the overall profiles are tilted.
 Note that the peak flux is shifted by $\approx 2,000$~\kmps .
 For intermediate angles (i.e. $-30^o, ~0^o,~30^o$), the profiles shift smoothly between the two extreme cases.

The observed profile of \qxp\ shows a flat-tilted (\feiiprofile) profile, that is highly
asymmetric. The tilt shifts the maximum flux towards the red,
suggesting that \qxp\ is observed from the south pole (i.e. the side opposite to the DDT point). The comparison
with Models 00, 03, and 09   (Fig. \ref{IR}, lower right plot) shows good
agreement between the observed spectrum and Model 03 seen from
an angle of $-90^o$ (Fig.  \ref{IR}, lower right).

\subsection{The overall NIR (and optical) spectrum}
\label{sect:overallIR}

The synthetic and observed spectra of \qxp\ are compared in Fig.~\ref{IR_spectrum}.
\newtextc{We use relative fluxes due to the absolute flux uncertainties in both the observation and model. In case of the observation,
the distance is uncertain by about $0.43$~mag (see \S 2), and no IR photometry are available to accurate   flux calibrate the spectrum. On the other hand,  in the case of the models the absolute IR flux may change by small adjustments in the $^{56}$Ni mass which may be needed to reproduce the early spectra, and in addition, would change the typical ratio between the second and third ionization-state.} 
The overall synthetic flux distribution seems to be somewhat steeper from the blue to the red compared to the observation, \newtextc{ but the agreement of line strengths of different ions is generally good.}
The contributions of individual transitions are identified in Table \ref{table.prediction}. The strongest and most of the medium strong features can be reproduced, namely at $1.04, 1.27, 1.55, 1.644, 1.8, 1.9, 2.07, 2.27$~\microns. The features are mostly produced by blends of singly and doubly ionized Fe and Co lines within $\approx \pm 2\% $ in wavelength of the peak given. \newtextc{The feature at 1.2 \microns can be attributed to fluorescence of [Fe II] emission by neutral Fe \citep{Wilk2018,Shingles2020}.}

The line profiles of the strong 1.27, 1.55~\microns  features are clearly asymmetric even in Model 00 because of blending between [\FeII]--[\FeIII] and [\CoII]--[\CoIII] lines (see Fig.~\ref{IR_spectrum2} and Table~\ref{table.prediction}). For example, the tilt of the $1.25$~\microns feature shows a steep blue and flat red decline produced mostly by multiple blends of
some 30+ transitions dominated by [\FeII], whereas the opposite is true for 
the $1.55$~\microns feature which is dominated by two [\CoIII] and [\FeII] transitions.  As shown by \citet{penney14} and \citet{Diamond2015}, the time evolution of the feature  at $1.55$~\microns can be understood by the
$^{56}$Co$  \rightarrow ^{56}$Fe  decay.  This, in principle, enables us to separate the effects of blends and  of the asymmetric envelope on the resulting profile. In practice, the separation of time and asymmetry effects may be challenging and requires good resolution and high S/N observations.

One  feature can be seen at $1.9$~\microns in the  synthetic spectra and tentatively seen in the observed spectra, although it is in the telluric region. In the model, it can be attributed to [\NiII] and previously has been identified in normal bright SNe~Ia by \citet{Friesen2017}.  In our models this feature is rather weak despite the significant amount of EC $\approx 0.08 M_\odot $ in the models because high densities in the central region suppress forbidden lines and the photosphere has not yet receded to the center (see \S~\ref{sect:Numerics}). The overall feature is blue-shifted and `stubby' without a red-shifted peak because, in our model, the EC distribution is rather spherical. As evident from Table \ref{table.prediction}, [Ni II] transitions contribute significantly to the emission, but as blends. Note that the strength of the feature depends on collisional de-excitation.

Some features in the synthetic spectra are present but too weak
compared to the observation, e.g. in the region at $1.4 $ and $2.15$~\microns.  From the models, emissions may be attributed to multiple
transitions of [\FeII] or, in the latter case, to [\NiII] and [\FeII]. In principle, higher ionization to  [\FeIII] may help because
this ion has lines at the corresponding wavelengths, in particular,
the [\FeIII] 3H-3G transitions at $2.144$ and $2.146$~\microns.  However, in
our models, this would decrease the $1.644$~\microns [\FeII] relative to
the Fe/Co blend at $1.55$~\microns. The discrepancy between models and
observations  points towards missing cross-sections as a potential problem: some 50\% of all [\FeII-\FeIII] and [\CoII-\CoIII] \newtextc{transitions} have non-measured cross-sections (see \S~\ref{sect:Numerics}), a problem not unique to our models \citep{Friesen2017}.
Another uncertainty may be the lack of collision rates which suppress
forbidden lines at higher excitation levels, or may boost allowed transitions which are known to be important to reproduce spectra in this wavelength range at earlier times in both normal bright and sub-luminous SNe~Ia \citep{Wheeler1998,HGFS99by02,Friesen2017}.  
 
The optical spectrum of
Models 00 and 03 is shown as an inset in Fig. \ref{IR_spectrum}.
It is dominated by singly and doubly ionized Fe/Co and C I, O I, and S I-II.
The very prominent feature is the [\CaII] doublet at $0.729,0.732$~\microns
blended with [\FeII], [\NiII] and [\CoII]. In our models, the feature
is expected to shoot up between 100 and 200 days after the explosion when the densities in
the corresponding Ca-rich layers (Fig.~\ref{models})
drop below the
critical density of our  atomic models (see \S~\ref{sect:Numerics}). [\CaII] is strong  in our  model due to  the
strong overlap of Ca and $^{56}$Ni produced during the detonation. The
total mass of  $^{56}$Ni produced is smaller by a factor of about two compared with normal-bright SNe~Ia which show only a very limited overlap between Ca and $^{56}$Ni and, thus, much weaker [Ca II]. 

The difference between the profile of the spherical and off-center
models suggests  [\CaII]  as a new diagnostic to probe asymmetries
(see inset of Fig. \ref{IR_spectrum}): 
In the spherical case, the feature is broad because the Ca exists in a
ring from $\approx -7,000 $ to $+7,000 $~\kmps. For an observer, the Ca-rich region spans a wide velocity range from red to blue-shifted segments (Fig. \ref{models}, upper figure). Strong asymmetries will produce narrow [\CaII] features, allowing the appearance of the blends mentioned above (Fig. \ref{IR}). The profile of Model 03
is similar in height, but much narrower than in  the spherical model because Ca is only strong in a specific segment of the envelope rather than a sphere (Fig. \ref{models}, lower middle plot).
 If seen from the north pole, the Ca-rich region comes towards the observer, but the red-shifted component is missing, and vice versa if seen from the south pole.  
The width of the  [\CaII] component  depends strongly on the direction to the observer. The projected velocities of Ca can be directly read off from Fig. \ref{models} (lower middle plot).
From $-90^o$ the contribution has a mean half width is $ \approx 2,500$~\kmps,  but goes down to $\approx 1,000$~\kmps if observed from the equator.

\section{Final Discussion}
\label{sect:DC}

Here, we want to put our detailed analysis of the IR spectrum obtained for \qxp\ into connection with the explosion and progenitor and discuss the results in context of various other leading scenarios.

\subsection{Delayed-Detonation Models with an Off-Center DDT}
\label{sect:DD}

 \qxp\ shows a highly asymmetric [\FeII] $1.644$~\microns NIR line profile.   This profile is reproduced using a full non-LTE 3D modeling code  for an  SN~Ia within the framework of off-center delayed-detonation models.

The distinguishing components of the NIR [\FeII] feature at $1.644$~\microns
are flat (with no curvature), but \feiiprofile profiles, which
includes a steeply declining component. In addition, there are
extended wings due to weak  blends of singly and doubly ionized Fe and
Co, respectively.  The extent of the hypotenuse of the \feiiprofile
profile does not depend on the location of the off-center detonation,
and can be used to determine the extent of the inner $^{56}$Ni-free region. The
tilt of the hypotenuse is an effective tool to constrain the
angle between the symmetry axis and the observer (see \S~\ref{Sect:lineprofiles}, and Figs. \ref{sect:models} $ \& $ \ref{IR}). 

The asymmetric profile of \qxp\ can be understood within the framework of delayed-detonation models with a high central density and a DDT triggered in the regime of distributed burning consistent with the Zeldovich mechanism. 
The DDT occurs on an already expanding background resulting in asymmetric abundance distributions because, for the same mass coordinate in the WD,  burning happens at different densities (see \S~\ref{sect:models}).

The high S/N spectrum of the  [\FeII] $1.644$~\microns feature enables us to  separate asymmetry from peculiar velocities (see \S~\ref{Sect:lineprofiles}).
We show that the shift of the peak flux of $\approx 2,000$~\kmps can be mostly attributed to the asymmetry in the explosion rather than a Doppler
shift reflecting the motion in the progenitor system, or in the host galaxy.
The offset of $\approx -500$ km~s$^{-1}$ is based on the steep red wing and the center of the short leg of the \feiiprofile profile. 
The latter offset is consistent with other NIR features pointing to a kinematic offset, i.e. orbital velocities.  

We showed that the overall observed and theoretical NIR spectra agree
within uncertainties expected from atomic physics (see
\S~\ref{sect:Numerics}). We verify that the off-center delayed-detonation model seen from the south pole can reproduce the spectra of \qxp .
It is shown that the strength of the features hardly depends on the off-centerness of the features (see Fig.~\ref{IR_spectrum}), but the profiles do (see Fig.~\ref{IR_spectrum2}.)

We weakly suggest the presence of the 1.9~\microns [\NiII] line due to stable $^{58}$Ni (\S~\ref{sect:Spectral}), which if verified  indicates high-density burning and therefore a high-mass WD progenitor. Moreover, a high density WD and the production of EC elements are  essential ingredients to boost the line asymmetry (see Figs. \ref{IR} \& \ref{IR_spectrum}, and \S~\ref{sect:overallIR}) and making \qxp\ a peculiar underluminous SN~Ia. Most SNe~Ia originate from lower $\rho_c$ (see \S~\ref{sect:motiv} \& \S~\ref{sect:Numerics}).

We also produced a synthetic, optical spectrum in which we identified
the [Ca  II] doublet at $0.7293/0.7326 $ $\mu$m 
 as an important diagnostic (see
\S\ref{sect:overallIR}). For $M_{\rm{Ch}}$ explosions, the strength of the
Ca feature is predicted to increase with decreasing luminosity and
increasing initial  density of the WD because it places the Ca right
into {the power source, the $^{56}$Ni region}. 
EC reduces the $^{56}$Ni  production in the high-density burning
regimes where Ca is destroyed. The result is a very strong [\CaII]
$0.73$~\microns feature predicted by our high-density, low luminosity
model. 
 
 The profile and width of [\CaII] are sensitive to asymmetries and the
 viewing angle. With the DDT being off-center, the optical \CaII\ is rather
 narrow  because asymmetry limits the projected velocity range to
 $\approx 1,000-2,000$~\kmps  compared to the spherical model which
 produces widths of $\sim 7,000$~\kmps  as discussed in \S
 \ref{sect:overallIR} (see Figs. \ref{models} $\&$ \ref{IR}, inset).

\subsubsection{\qxp in context of other peculiar and sub-luminous SNe Ia}

 
 Within the delayed-detonation scenario, the underluminous \qxp\ mainly differs 
 by the high central density $\rho_{\rm c}$ from the transitional SN~Ia SN~2007on with $\rho_{\rm c} $ of $10^9$g~cm$^{-3}$. The higher $\rho_{\rm c}$ of \qxp\ lowers the peak brightness and
 shifts the luminosity decay rate from a transitional SN~Ia to a slower decliner (\S~\ref{sect:LCP}, Tab. \ref{table0})
 and into the regime of the luminosity decline rates commonly occupied by Ca-rich transients.
 Moreover, we want to note that the $\rho_{\rm c}$ of \qxp is relatively close
 to the  accretion-induced collapse limit, similar to the peculiar sub-luminous
 SN~2016hnk \citep{Galbany19} discussed in \S~\ref{sect:motiv}.

Note that spectropolarimetry data have been obtained for many normal SNe~Ia and, most of them,
show high line polarization in Si II consistent with off-center delayed-detonation models \citep{2006NewAR..50..470H,Cikota19}.  
What seems to make \qxp peculiar and amplifies the line asymmetry is the  high $\rho_{\rm c}$ in an underluminous SN~Ia (\S~\ref{sect:models}).
   
\subsubsection{The progenitor system of \qxp }

 The offset in velocity relative to the rest frame in the peak flux of a particular feature is commonly used to estimate the orbital motion within the progenitor system, and the motion of the system in the galaxy.  High velocities are taken as evidence for DD progenitor systems \citep{Diamond2015,Diamond2018,maeda11,2018MNRAS.477.3567M}. Blends are
 the obvious problem for most of the strong optical and IR features (\S~\ref{sect:overallIR}), and intrinsic
 asymmetries in the density and abundance distribution pose another difficulty.
  
 As shown in \S~\ref{sect:obs}, the peak flux of \qxp\ of the unblended \feiiprofile profile is $\approx 2,000$ \kmps. 
 The center of the central part of the \feiiprofile profile provides a measure of the offset of the system, whereas the peak flux is a measure of the intrinsic asymmetry in the chemical distribution of the SN ejecta. Note that the central offset does not depend on the viewing angle of the observer (\S~\ref{sect:obs} \& Fig. \ref{IR}) whereas, obviously, the flux-averaged center of the line flux would produce a systematic shift. We find an offset in the profile and spectra, $v_{\rm{off}} \approx 500$ \kmps (\S~\ref{Sect:lineprofiles}),  with respect to the redshift measured by the Doppler shift derived from
Arecibo H I 21 cm measurements of the entire galaxy. According to \citet{Schneider1990}, the observations show a profile with widths of $W50 = 77 \pm 8$\kmps and $W20 = 127 \pm 12 $\kmps. With \qxp\ far away from the galactic center, adding an  uncertainty in $v_{\rm off}=500 \pm 127$ \kms. Still, the low value of $v_{\rm off}$ may suggest a SD-system although close as the progenitor of \qxp.

\subsection{Alternative Explosion Scenarios}
\label{sect:Dalternatives}
Here, we put our work in context of other leading explosions scenarios which have been  discussed in \S~1.

\subsubsection{He-triggered Detonations}
\label{sect:DHeD}
In light of the large parameter range yet to be explored for this class,
we cannot exclude these scenarios. Instead, we limit our discussion to models published.

 The small velocity offset of the line  profiles of \qxp (\S~\ref{IR} \& \S~\ref{sect:DD})
 is not to be expected for a system originating with a compact donor unless
 seen from the polar direction. 


In the He-triggered detonation scenario, the progenitors of normal-bright SNe~Ia are on the high mass end, $\ga 1.1$~\Msun  and can reach the regime of EC, but this is not the case for low-mass WD progenitors  commonly attributed to transitional SNe~Ia  dimmer than  $-18$~mag,  which have progenitor WD masses less than 1 \msol \citep{Scalzo19,Polin19,Gronow:2021b}. While \qxp is not a transitional SNe~Ia, it is low luminosity and therefore would fall into the sub 1 \msol case in the helium detonation scenario.
The possible detection of [Ni II] at 1.9 $\mu$m would be inconsistent with this class of models.

Most models of He-triggered detonations have assumed that the inner
detonation occurs at the center \citep{hk96,Blondin15,Shen2015,Polin19,Shen2021}, with rather spherical distributions of the iron-group elements that, for under-luminous SNe~Ia, are located in the central region.
In reality, however, the detonation of the CO core may well be triggered off-center \citep{livne95,livne99}.
In principle, such off-center carbon detonations may produce asymmetries. However, those are likely to be insufficient to produce the 
spectrum and highly asymmetric line profile of \qxp because those require a significant amount of $^{56}$Ni being in chemically asymmetric layers. 
In pure detonations, the unburned material is not affected by the energy input at other places
because the front propagates as a weak detonation, i.e. close to the speed 
of sound. In a spherical WD, the burning conditions of the material do not depend on the point where the
detonation in the CO-core is triggered and whether it is burned later,
the mechanism which works well in off-center delayed-detonation models. 
To be successful, He-triggered models may require very asymmetric initial
density distributions close to the center.

\subsubsection{Colliding WDs}
\label{sect:DCWD}

 WD mergers 
 through direct WD collisions (\S~1) are prime candidates for explaining the highly asymmetric \feiiprofile at  $1.644$ \microns of \qxp.
 The small offset of the line profile (\S~\ref{IR} \& \ref{sect:DD}) is consistent with this scenario (\S~1).
 The tentatively identified forbidden [\NiII] line at 1.9 \microns (\S~\ref{sect:Spectral}) would require that one of the exploding  WDs has a high central density (and mass $> 1.1 M_\odot$) to allow for the detonation front to compress the  central region and produce EC elements \citep{Hoeflich2018NUC}. The Ni line would not be a show stopper.\footnote{Note that we did not test the influence of the high density on the overall spectra nor the effect of high-density equation of state on the hydrodynamical solution of colliding WDs.} 
 
 \citet{Dong2015} presented a detailed study of line profiles in colliding WDs. 
 The $^{56}$Ni is off-center in each of the two exploding WDs. The profiles show `double-horned' features that depend on the parameters and aspect angle and can be rather asymmetric in the strengths of the two peaks or even merge. \citet{Dong2015} noted that, at low resolution, the double-peaked profile  may appear as a flat profile, or 'pot belly` without the double-peak if seen from directions orthogonal to the line of collision. However, the central component of the \qxp profile spans a large velocity range of $\approx 5,000$ \kmps (Figs. \ref{fig:data} \& \ref{IR}) that is comparable to the separation to be expected for the two $^{56}$Ni regions within this model. Likely, an observer has to view the explosion along the axis of the collision. In contrast to the prediction \citep{Dong2015}, the observed spectrum has  medium-resolution and high S/N and does not show two broad horns in the $1.644$ \microns feature.
 
 Note that one strong IR feature at $1.55$ \microns shows double-horns (Fig. \ref{fig:data}). However, this feature is dominated by two separate transitions, by  [\CoIII] and by [\FeII], and not indicative of asymmetry (see \S~\ref{sect:Numerics}, \S~~\ref{sect:Spectral}, and Table~\ref{table.prediction}). Because the profile is directly linked
 to the main decay products, the double feature should be 'generic` to all scenarios.
 
 \citet{Dong2015} produced profiles for direct collisions of two WDs of similar masses, which results in a flat top rather than a tilt with the flux of the blue and red edges differing by a factor of $\approx 2$. In principle, one can change the mass ratio of the two colliding WDs. Because two WDs with masses $M_{1}$ and $M_{2}$ are hydrostatic, they can be described by polytropes with a radius ratio $R_1/R_2 \approx (M_2/M_1)^{1/3} $.
 If we assume that the center of the WDs are ignited and using the  $^{56}$Ni
 production in pure detonations of a CO WD \citep{hk96,Shen14,Blondin17}, $M_1/M_2 $ should be about $0.8$ with $R_1/R_2 \approx 1.07$. This small a difference in the radii is too little to invalidate \citep{Dong2015} examples, and to widen the line profile sufficiently to allow observing closer to the axis of impact to avoid
 double-horns.  Both the lack of 'double-horns' and the linear central part of the profile \qxp\ are not predicted for colliding WDs, and
 seem to be inconsistent with the observation.
 
 \subsubsection{Dynamical merging of two orbiting WDs}
 \label{Dmerger}
The post-explosion structures of dynamical mergers of two sub-M$_{\rm {Ch}}$ WDs depend sensitively on the mass-ratio
between the WDs (\S~1). 
Typically, the dynamics involves a quasi-hydrostatic state
of a rapidly rotating configuration and leads to axial symmetry in both  density and abundance distribution
of rapidly expanding envelopes (see e.g. \citet{Garcia-BerroHB}, 
their Fig. 1). The resulting $^{56}$Ni distributions
are central, or in torus-like regions. The resulting line-profiles are close to Gaussian or double-horned, respectively, \citep{gerardy03du04} unlike  the \feiiprofile profile observed in \qxp.
Post explosion, the offset in velocity is small, consistent with the offset observed in \qxp . 

A variant of dynamical mergers are the so-called violent mergers as a result of two WDs with very different masses \citep{pakmor12}. The interaction between the two WDs preluding or during the merger creates a hotspot on
the surface of the primary, more massive, CO-WD that triggers a detonation in the massive CO-WD. The result of the explosion is a one-sided,  off-center `banana'-shaped $^{56}$Ni distribution (see Fig. 1 in \citet{PakmorHB}), very similar to the Ca distribution in the off-center DDT models (Fig.\ref{models}). No line profile has been published, but it can be inferred from the approach in \S~\ref{Sect:lineprofiles} \& \S~\ref{sect:overallIR}.
Seen from the north and south pole, the 
$1.644$ \microns feature should show a narrow blue or redshifted profile, rather than a peaked and asymmetric profile similar to the [CaII] in off-center delayed-detonations (Fig. \ref{IR_spectrum}).
For low inclinations, we expect peaked, unshifted profiles, and with smooth transition to broader profiles with a shift of the center, if seen from intermediate directions. Such profiles are inconsistent with the observation. The main difference to our off-center delayed-detonation models is the lack of an isotropic $^{56}$Ni component produced during the deflagration phase. Also, one may expect a large offset of the line profile. 

As above, none of the dynamical mergers are likely consistent with the presence of EC elements,
namely the tentatively identified Ni feature at 1.9 $\mu$m.

\subsection{Spectropolarimetry} 
\label{sect:specpol}
Snapshots of NIR nebular line profiles in combination with detailed, 3D  non-LTE models probe asymmetry in SNe~Ia of the inner region of the ejecta, and are highly complementary to spectropolarimetry  \citep{h95pol,2001PASP..113.1178L} programs such as SPECPOL at the VLT.  Although expensive, polarization is a multipurpose tool to measure asymmetry of the outer QSE layers and probes SNe~Ia physics during the photospheric phase \citep{2003ApJ...591.1110W,Leonard2005,2006NewAR..50..470H,Kasen2006,wang08,2009A&A...508..229P,Tanaka2010,2013MNRAS.433L..20M,Bulla16,2019MNRAS.490..578C,Yang2020}. It may be noted that the highest continuum polarization ($\approx 0.5\%$), which is produced by Thomson scattering, has  been observed in underluminous SNe~Ia such as SN1999by and SN2005ke. There, the axis of symmetry is well-defined and does not change with time. Moreover, line polarization over the optical Si/S lines during the photospheric phase of many SNe~Ia indicate large scale asymmetries in QSE abundances with a well-defined axis of symmetry(see \S~\ref{sect:motiv}) which is consistent with 
off-center DDTs \citep{howell01,patat12,2006NewAR..50..470H}. For the underluminous \qxp, the nebular spectra probe directly the orientation and distribution of  $^{56}$Ni. Thus, spectropolarimetry and late-time snapshots
are highly complementary tools to decipher
the full 3D structure of thermonuclear SNe.

\subsection{Model Limitations}
\label{sect:limits}
 Finally, we also want to mention the limitations of our analysis.
 In particular, the separation between ISM  extinction and intrinsic color of \qxp\ is beyond the scope of this paper, and will be a topic of future work (S.~Bose et al., in preparation). The model parameters have not been tuned and full, time-dependent 3D non-LTE simulations may be required for a detailed analysis of flux spectra during the photospheric phase. The combination of spectropolarimetry and late-time NIR snapshots is a promising path to measure the 3D structure of SNe.  As a next step, we plan time-independent simulations for the photospheric phase data based on the asymmetries derived in this paper. Here we focused on one particular model, an off-center delayed-detonation model,  and
presented an argument of why it is favored.

\section{Conclusions}
\label{sect:concl}
We have shown the wealth of information that can be
extracted from a single late-phase NIR spectrum with high S/N.
We present a Keck (+ NIRES) NIR-nebular phase spectrum of \qxp, at
+191~d relative to rest-frame days past the epoch of $B$-band
maximum. We identify \qxp\ as  peculiar, underluminous SN~Ia, 
and from our modeling efforts find that it could originate from the
explosion of a high-density, Chandrasekhar mass ($M_{\rm{Ch}}$) WD rather
close to an Accretion Induced Collapse (AIC), similar but somewhat brighter than SN~2016hnk.
Such high density can be obtained in a progenitor system with low accretion rates onto the exploding WD. The small offset velocity $v_{\rm {off}}$ may favor small orbital motion in an SD system (\S~\ref{sect:DC}).

We showed that the $1.644$~\microns feature is  a unique spectroscopic tool obtainable from the ground that cannot be substituted by other, comparably strong features in the optical and NIR which are blends
(Figs.~\ref{IR_spectrum}--\ref{IR_spectrum2}, Table \ref{table.prediction}, Tables A1-A4). It allows to decipher the 3D geometry of the distribution of iron-group elements, very complementary to spectropolarimetry
during the photospheric phase as discussed above.
Additionally, the MIR offers many strong, minimally blended features
which have been used to verify the analyses of the $1.644$~\microns
[Fe II] feature, as was the case for
 SN~2004du observed by the Spitzer Space Telescope and for SN~2014J
 observed by the  Grand Telescopio Canarias
 \citep{gerardy03du04,Telesco2015}, although the resolution required
 for detailed MIR analyzes will only be obtained with JWST
 \citep{2021jwst.prop.2114A}. 
JWST
will open the MIR with many minimally blended lines,  which will allow for
detailed measurements of individual abundance distributions of many
elements and isotopes, and without  water vapor and other atmospheric
interference \citep{2021jwst.prop.2114A}. 
However, targeted high S/N snapshots for a significant number of 
SNe~Ia with NIRIS/KECK or VLT/E-ELT can be obtained to understand the apparent and intrinsic diversity of SN~Ia progenitors and the way in which nature drives their explosions \footnote{As discussed above, the profiles produced are different for off-center DDTs (tilted), head-on collisions (double-horned and tilted), mergers (double horned, likely equal), and HeD (rather symmetric) if observed off the equator. However,  for a statistical sample, the number required may be as large as 100 spectra if all explosion channels are realized.}.

A new generation of 3D non-LTE simulations is looming, but the
numerical efficiency needs to be optimized to be on par with spherical
simulations, and gaps in the atomic cross-sections need to be filled. 

\acknowledgments
\newtextc{We thank the referee for the careful reading, useful comments and pointing  out the initial problem with the 1.2 \microns  feature.}
PH, AF, BH, DC, and EYH acknowledge support by the National Science
Foundation (NSF) grant AST-1715133. EB was supported in part by NASA
grant 80NSSC20K0538. CA and BJS are supported by NASA grant 80NSSC19K1717 and NSF grants AST-1920392 and AST-1911074.
MS is supported by grants from the Villum FONDEN
(28021) and the Independent Research Fund Denmark (8021-00170B). MAT
acknowledges support from the DOE CSGF through grant DE-SC0019323. 
L.G. acknowledges financial support from the Spanish Ministry of Science, Innovation and Universities (MICIU) under the 2019 Ram\'on y Cajal program RYC2019-027683 and from the Spanish MICIU project PID2020-115253GA-I00. CRB, EYH, MMP, NBS and PH are supported by the NSF grant AST-1613472. We thank Prof. Valenti for his support to SD and helpful discussions.

\bigskip
\facility{The data presented herein were obtained at the W. M. Keck Observatory, which is operated as a scientific partnership among the California Institute of Technology, the University of California and the National Aeronautics and Space Administration. The Observatory was made possible by the generous financial support of the W. M. Keck Foundation. This research has made use of the Keck Observatory Archive (KOA), which is operated by the W. M. Keck Observatory and the NASA Exoplanet Science Institute (NExScI), under contract with the National Aeronautics and Space Administration. 
The authors  recognize and acknowledge the  significant cultural role and reverence that the summit of Maunakea has within the indigenous Hawaiian community.  We are  fortunate to have the opportunity to conduct observations from this majestic mountain.}
\facility{The simulations have been performed on the computer cluster
  of the astro-group at Florida State University.
}

\noindent
\software{
 \hydra\ \citep{hoeflich2003hydra,Hoeflich:hydro:2009,Hoeflich2017}, SNooPy \citep{Burns11} and OpenDx an open-sourced graphics package by IBM. 
The {\it IDL} packages {\it Spextools} and  {\it Xtellcorr} in version 5.0.2 were used for the data reduction, flux calibration and correction of telluric feature \citep{spextool}.
} 
\clearpage
\bibliographystyle{aasjournal}
\restartappendixnumbering

\begin{appendix}
\section
{Ionization, $\gamma $-rays and non-thermal leptons}
\label{appendix:A}

Detailed simulations of nebula spectra  and ionization balances have been 
pioneered by \citet{Axelrod1980}. Significant advances have been achieved by \citet{KF92} by treating the non-thermal excitation and ionization  using the Spencer-Fano equations \citep{Spencer54}. 
Ionization balances during the nebular phase have been widely studied in 
the literature, often with discrepant results \citep{Liu98,Ruiz-Lapuente1992}. Advances in computational physics over the next decades led to increasingly sophisticated codes for nebular spectra \citep{2015ApJ...814L...2F,2017ApJ...845..176B,Shingles2020,Wilk2018,Wilk2020} sometimes based on or extensions of existing non-LTE codes (e.g. \citep{hillier90,Kasen2006,Sim10}) or, in this work, HYDRA \citep{Hoeflich:hydro:2009,Hristov2021} (see \S 4.1). The various codes are still evolving and differ, among others, by the treatment of non-thermal ionization and excitation by $\gamma$-rays, radiation and positron transport including magnetic fields,  nuclear and atomic reactions involved and whether the cascading down of non-thermal leptons is based on experiments, theory or treated by Monte-Carlo.

A  discussion of the physical processes involved and evaluation of the differences of codes is beyond the scope of this paper. For a detailed discussion of physical processes, we would like to refer to \citet{Shingles2020}. In Sect. \ref{sect:Numerics}, the possible implications of the physical assumptions have been presented.
Because the results presented in this paper hardly depend on the details below, we want to summarize our approach used in this particular work to calculate the nebular spectrum of our model for \qxp at about 210 days after the explosion.

We assume the radioactive $^{56}$Ni decay chain, with positrons originating from the $\beta^+$ decay of $^{56}Co$ and pair-production. The energy deposition by hard $\gamma $-rays and non-thermal electrons and positrons enters the rate equations via non-thermal ionization balanced by the recombination processes similar to \citet{KF92}, under the assumption that the local nuclear energy input per time is balanced by the local flux. The basic set of equations is given in the appendix B of \cite{Hristov2021}.

 For X- and $\gamma$-rays and non-thermal leptons, i.e. positrons and electrons, a Monte Carlo Scheme is used to calculate  the interaction with the matter by Compton scattering on electrons, scattering on nuclei in the plasma, annihilation $e^+e^- \rightarrow 2\gamma$  or $e^+e^- \rightarrow 3\gamma$ for para- and ortho-positronium, and by the reverse processes. In addition, we employ bound-free absorption of the inner-shell electrons in the X-rays \citep{Berger1998,hkm92,Hoeflich:gamma2002}\footnote{Version from 2010, XCOM: Photon Cross Section Data Base,https://www.nist.gov/pml/xcom-photon-cross-sections-database}.  In this paper, we assume that the probability of an interaction with a specific ion is proportional to the abundance and number of bound and free electrons. The exception is $^{56}$Co for which we add the interaction of the primary $\beta^+ $ positron using the electron density in its electron shell. The energy loss of the non-thermal leptons is reduced by the ionization energy of the level undergoing interaction. The corresponding level is filled by a series of radiative transitions corresponding to the energy difference between levels between non-valence electrons and Auger transitions, using the probability of multiple electron emission by \citet{Kaastra1993}. Because of our limitation to four ionization states, we treat
 multiple-ionizations in a single ion as multiple single-ionizaton processes in the rates.  
 The interaction with atoms and free electrons enters into the thermal bath of the plasma and produce directly additional non-thermal electrons. The low-energy cutoff for non-thermal electrons and positrons is given by the ionization energy of the ions included in the detailed network and the binding energy of positronium, respectively. If the energy of a lepton is less than the ionization energy, it is treated as excitation rate corresponding to the level populations 
 and bound-bound cross-sections.
 For computational efficiency and because the number of non-thermal leptons increases with decreasing energy, the non-thermal leptons are grouped into 100 energy bins to follow the cascade. 
 
 For the recombination, we use both stimulated and spontaneous emission to both the ground and excited states. Note that, in particular in the UV (and somewhat in the $U$ and $B$ bands), the envelope is partially optically thick for many allowed bound-bound and bound-free transitions from the lower levels of each ion resulting in an effective redistribution from the UV to longer wavelengths by incomplete Rosseland cycles which  couple higher to lower ionization states. Because the number of UV transitions increases with the ionization level, the UV field effectively limits the ionization to higher ions. 
  \begin{figure*}[ht]
    \centering
    \includegraphics[width=0.9\textwidth]{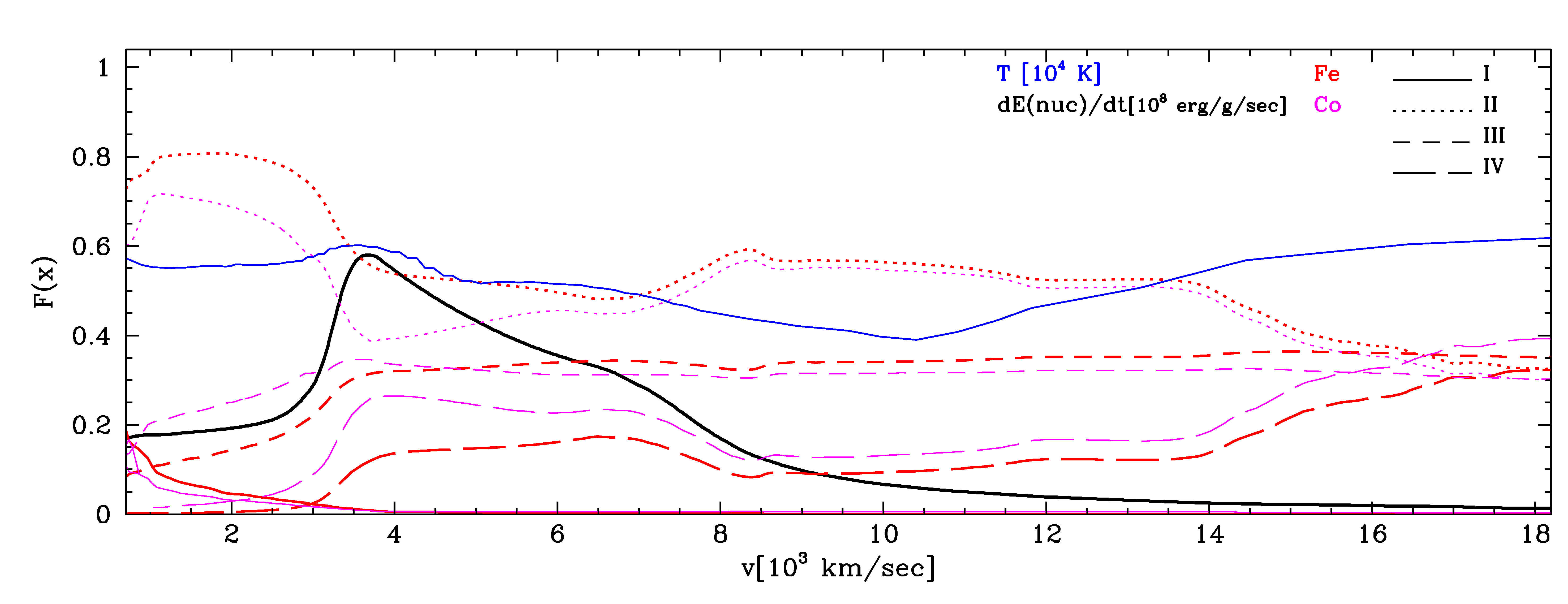} 
    \caption{Angle averaged ionization levels I-IV of Fe and Co as representative of iron-group elements, specific energy input by $\gamma $-rays and non-thermal leptons, and temperature T as a function of expansion velocity of the underluminous model used
    for \qxp.}
\label{fig:ion}
\end{figure*}

 For illustration and as reference, the ionization level for Fe and Co is shown in Fig. \ref{fig:ion} as representatives for iron-group elements. 
 Intermediate mass elements, where present in the envelope, show a similar pattern but shifted to larger II/III/IV ratios.   Because of the spherical density distribution, deviations from spherical symmetry remain rather small.  
 In the line-forming region, the temperatures are 4000-5500 K with a peak coinciding with the maximum energy deposition. Unlike normal-bright SNe~Ia of $M_{Ch}$ models with double ionized ions dominating at about day 210 \citep{Wilk2018,Shingles2020}, our subluminous SNe~Ia shows the second ionization state dominating in the line-forming region.  The overall lower 
 dominant ionization stage is a result of the reduced $^{56}$Ni mass, by a factor of $\approx 2 ... 3$, and because the average $^{56}$Ni is shifted to the  inner, higher density regions (Fig. 2) boosting the recombination rate. Already by day 210, we see neutral iron group elements in the center. In our models, the rise of the ionization stage to III and IV at high velocities is caused by the low densities reducing the recombination rate, and the lower UV optical depth 'shutting off' of the incomplete Rosseland cycle towards lower ions.
 Note that the ionization state shows only moderate variations in the line-forming region of the $1.644$ \mum ~ feature lending additional stability to analysis of line profiles in this paper as discussed in Sect. 4.1.
 
\clearpage

\section{Optical Line Lists}
\label{appendix B}
In Tables~B1-B4 we present line lists
for relevant optical forbidden lines.
\begin{table*}
\begin{center}
    \caption{Same as Table \ref{table.prediction} but for synthetic spectra in the optical (Part 1).}
    \begin{tabular}{|rr|rr|rr|rr|rr|}
    \hline
    S~~~$\lambda $[\microns] &Ion  &   
    S~~~$\lambda $[\microns] &Ion  &   
    S~~~$\lambda $[\microns] &Ion  &   
    S~~~$\lambda $[\microns] &Ion  &   
    S~~~$\lambda $[\microns] &Ion  \\
\hline
          0.3052 &    [Co II]  & 
  **      0.3077 &    [Ni II]  & 
   *      0.3083 &    [Co II]  & 
          0.3160 &    [Co II]  & 
          0.3174 &    [Co II]  \\
  **      0.3176 &    [Fe II]  & 
   *      0.3193 &    [Co II]  & 
          0.3202 &    [Co II]  & 
          0.3205 &    [Co II]  & 
   *      0.3216 &    [Fe II]  \\
          0.3219 &    [Co II]  & 
          0.3223 &    [Co II]  & 
  **      0.3224 &    [Ni II]  & 
   *      0.3225 &    [Fe II]  & 
          0.3241 &    [Fe III] \\
   *      0.3244 &    [Co II]  & 
   *      0.3245 &    [Fe II]  & 
  **      0.3255 &    [Fe II]  & 
          0.3259 &    [Co II]  & 
   *      0.3276 &    [Fe II]  \\
   *      0.3278 &    [Fe II]  & 
   *      0.3279 &    [Fe II]  & 
   *      0.3290 &    [Fe II]  & 
   *      0.3291 &    [Fe II]  & 
          0.3296 &    [Co II]  \\
          0.3306 &    [Co II]  & 
   *      0.3306 &    [Co II]  & 
          0.3314 &    [Co II]  & 
   *      0.3319 &    [Fe II]  & 
          0.3326 &    [Co II]  \\
          0.3326 &    [Co II]  & 
          0.3364 &    [Co II]  & 
   *      0.3366 &    [Co II]  & 
   *      0.3377 &    [Co II]  & 
  **      0.3377 &    [Fe II]  \\
          0.3378 &    [Co II]  & 
  **      0.3379 &    [Ni II]  & 
   *      0.3382 &    [Fe II]  & 
   *      0.3382 &    [Co II]  & 
          0.3383 &    [Co II]  \\
          0.3383 &    [Co II]  & 
  **      0.3388 &    [Fe II]  & 
   *      0.3402 &    [Co II]  & 
   *      0.3404 &    [Fe II]  & 
   *      0.3421 &    [Co II]  \\
   *      0.3429 &    [Fe II]  & 
          0.3434 &    [Co II]  & 
  **      0.3440 &    [Ni II]  & 
  **      0.3442 &    [Fe II]  & 
   *      0.3451 &    [Fe II]  \\
  **      0.3453 &    [Fe II]  & 
  **      0.3456 &    [Fe II]  & 
   *      0.3485 &    [Fe II]  & 
   *      0.3491 &    [Fe II]  & 
  **      0.3503 &    [Fe II]  \\
  **      0.3505 &    [Fe II]  & 
  **      0.3505 &    [Fe II]  & 
   *      0.3507 &    [Fe II]  & 
   *      0.3529 &    [Fe II]  & 
          0.3529 &    [Co II]  \\
   *      0.3531 &    [Co II]  & 
   *      0.3534 &    [Fe II]  & 
   *      0.3537 &    [Fe II]  & 
  **      0.3540 &    [Fe II]  & 
  **      0.3540 &    [Fe II]  \\
   *      0.3557 &    [Co II]  & 
          0.3560 &    [Ni II]  & 
  **      0.3560 &    [Ni II]  & 
          0.3565 &    [Co II]  & 
   *      0.3577 &    [Fe II]  \\
   *      0.3580 &    [Ni II]  & 
   *      0.3581 &    [Fe II]  & 
          0.3583 &    [Co II]  & 
   *      0.3584 &    [Co II]  & 
   *      0.3589 &    [Fe II]  \\
   *      0.3599 &    [Co II]  & 
   *      0.3606 &    [Fe II]  & 
   *      0.3614 &    [Co II]  & 
   *      0.3621 &    [Co II]  & 
   *      0.3627 &    [Fe II]  \\
          0.3628 &    [Ni II]  & 
  **      0.3628 &    [Ni II]  & 
   *      0.3630 &    [Fe II]  & 
          0.3635 &    [Co II]  & 
  **      0.3638 &    [Co II]  \\
   *      0.3640 &    [Co II]  & 
   *      0.3641 &    [Co II]  & 
   *      0.3643 &    [Fe II]  & 
   *      0.3654 &    [Co II]  & 
   *      0.3666 &    [Fe II]  \\
   *      0.3679 &    [Co II]  & 
  **      0.3682 &    [Co II]  & 
 ***      0.3689 &    [Co II]  & 
   *      0.3710 &    [Co II]  & 
   *      0.3727 &    [O II]   \\
   *      0.3730 &    [O II]   & 
          0.3754 &    [Co II]  & 
   *      0.3754 &    [Co II]  & 
  **      0.3769 &    [Co II]  & 
  **      0.3771 &    [Co II]  \\
   *      0.3771 &    [Co II]  & 
          0.3799 &    [Co II]  & 
  **      0.3802 &    [Co II]  & 
   *      0.3845 &    [Ni II]  & 
   *      0.3848 &    [Co II]  \\
   *      0.3863 &    [Co II]  & 
   *      0.3977 &    [Co II]  & 
   *      0.3977 &    [Co II]  & 
   *      0.3981 &    [Co II]  & 
   *      0.3994 &    [Ni II]  \\
   *      0.4018 &    [Co II]  & 
          0.4026 &    [Ni II]  & 
   *      0.4034 &    [Ni II]  & 
  **      0.4070 &    [S II]   & 
  **      0.4078 &    [S II]   \\
   *      0.4083 &    [Fe II]  & 
   *      0.4085 &    [Fe II]  & 
  **      0.4104 &    [Co II]  & 
   *      0.4112 &    [Co II]  & 
  **      0.4116 &    [Fe II]  \\
  **      0.4121 &    [Co II]  & 
          0.4126 &    [Co II]  & 
   *      0.4133 &    [Co II]  & 
   *      0.4144 &    [Ni II]  & 
   *      0.4146 &    [Co II]  \\
   *      0.4148 &    [Ni II]  & 
   *      0.4150 &    [Co II]  & 
   *      0.4150 &    [Fe II]  & 
 ***      0.4154 &    [Co II]  &
   *      0.4156 &    [Co II]  \\ 
   *      0.4159 &    [Fe II]  & 
   *      0.4171 &    [Co II]  & 
   *      0.4173 &    [Co II]  & 
   *      0.4178 &    [Co II]  &
  **      0.4178 &    [Fe II]  \\ 
   *      0.4180 &    [Fe II]  & 
          0.4189 &    [Co II]  & 
   *      0.4199 &    [Fe II]  & 
   *      0.4202 &    [Ni II]  &
  **      0.4212 &    [Fe II]  \\ 
   *      0.4216 &    [Co II]  & 
   *      0.4218 &    [Co II]  & 
          0.4225 &    [Co II]  & 
   *      0.4233 &    [Fe II]  &
   *      0.4236 &    [Fe II]  \\ 
 ***      0.4245 &    [Fe II]  & 
 ***      0.4246 &    [Fe II]  & 
  **      0.4246 &    [Co II]  & 
   *      0.4250 &    [Ni II]  & 
   *      0.4253 &    [Fe II]  \\ 
  **      0.4263 &    [Co II]  &
   *      0.4267 &    [Co III] & 
   *      0.4268 &    [Fe II]  & 
   *      0.4274 &    [Co II]  &
 ***      0.4278 &    [Fe II]  \\
   *      0.4281 &    [Fe II]  & 
   *      0.4286 &    [Ni II]  & 
 ***      0.4289 &    [Fe II]  & 
  **      0.4289 &    [Co II]  & 
   *      0.4294 &    [Co II]  \\
   *      0.4295 &    [Ni II]  & 
  **      0.4307 &    [Fe II]  & 
          0.4312 &    [Ni II]  & 
          0.4313 &    [Co I]   & 
          0.4316 &    [Ni II]  \\
   *      0.4317 &    [Co II]  & 
 ***      0.4321 &    [Fe II]  & 
  **      0.4324 &    [Co II]  & 
   *      0.4327 &    [Ni II]  & 
          0.4327 &    [Co II]  \\
   *      0.4331 &    [Fe II]  & 
          0.4337 &    [Co III] & 
          0.4345 &    [Co II]  & 
 ***      0.4348 &    [Fe II]  & 
   *      0.4353 &    [Co II]  \\
   *      0.4353 &    [Fe II]  & 
 ***      0.4354 &    [Fe II]  & 
   *      0.4357 &    [Fe II]  & 
 ***      0.4360 &    [Fe II]  & 
 ***      0.4361 &    [Fe II]  \\
   *      0.4362 &    [Co II]  & 
   *      0.4362 &    [Co II]  & 
  **      0.4364 &    [O III]  & 
   *      0.4365 &    [Co II]  & 
 ***      0.4374 &    [Fe II]  \\
          0.4377 &    [Co II]  & 
   *      0.4383 &    [Co II]  & 
  **      0.4384 &    [Fe II]  & 
   *      0.4384 &    [Fe II]  & 
  **      0.4389 &    [Co III] \\
          0.4401 &    [Co III] & 
   *      0.4404 &    [Fe II]  & 
   *      0.4411 &    [Co II]  & 
   *      0.4411 &    [Fe II]  & 
   *      0.4412 &    [Co II]  \\
 ***      0.4415 &    [Fe II]  & 
 ***      0.4417 &    [Fe II]  & 
   *      0.4418 &    [Co II]  & 
   *      0.4426 &    [Co III] & 
  **      0.4434 &    [Fe II]  \\
          0.4436 &    [Co II]  & 
   *      0.4438 &    [Co II]  & 
   *      0.4449 &    [Co II]  & 
          0.4449 &    [Co I]   & 
          0.4451 &    [Co II]  \\
 ***      0.4453 &    [Fe II]  & 
 ***      0.4459 &    [Fe II]  & 
   *      0.4461 &    [Co II]  & 
          0.4463 &    [Ni II]  & 
          0.4467 &    [Ni II]  \\
   *      0.4468 &    [Co II]  & 
   *      0.4470 &    [Fe II]  & 
  **      0.4470 &    [Co III] & 
  **      0.4472 &    [Fe II]  & 
 ***      0.4476 &    [Fe II]  \\
          0.4477 &    [Co I]   & 
   *      0.4480 &    [Fe II]  & 
          0.4487 &    [Ni II]  & 
  **      0.4487 &    [Ni II]  & 
  **      0.4490 &    [Fe II]  \\
 **      0.4494 &    [Fe II]  & 
 **      0.4501 &    [Co III] & 
         0.4501 &    [Co III] & 
 **      0.4509 &    [S I]    & 
 **      0.4511 &    [Fe II]  \\
 **      0.4516 &    [Fe II]  & 
***      0.4524 &    [Ni I]   & 
         0.4524 &    [Ni I]   & 
         0.4529 &    [Co II]  & 
 **      0.4530 &    [Fe II]  \\
         0.4530 &    [Co II]  & 
  *      0.4534 &    [Fe II]  & 
 **      0.4543 &    [Co II]  & 
         0.4549 &    [Co III] & 
         0.4562 &    [Co I]   \\
\hline
\end{tabular}
\end{center}
\label{tab:table2a}
\end{table*}

\begin{table*}
\begin{center}
    \caption{Same as Table \ref{table.prediction} but for synthetic spectra in the optical (Part 2).}
    \begin{tabular}{|rr|rr|rr|rr|rr|}
    \hline
        S~~~$\lambda $[\microns] &Ion  &   
    S~~~$\lambda $[\microns] &Ion  &   
    S~~~$\lambda $[\microns] &Ion  &   
    S~~~$\lambda $[\microns] &Ion  &   
    S~~~$\lambda $[\microns] &Ion  \\
\hline
          0.4565 &    [Co II]  & 
   *      0.4567 &    [Ni I]   & 
          0.4569 &    [Co III] & 
          0.4571 &    [Co III] & 
          0.4575 &    [Ni II]  \\
   *      0.4578 &    [Fe II]  & 
          0.4583 &    [Co III] & 
  **      0.4590 &    [S I]    & 
   *      0.4599 &    [Co I]   & 
          0.4608 &    [Fe III] \\
          0.4613 &    [Co II]  & 
          0.4617 &    [Co I]   & 
   *      0.4623 &    [C I]    & 
 ***      0.4624 &    [Co II]  & 
   *      0.4628 &    [Co III] \\
          0.4628 &    [Co III] & 
          0.4629 &    [C I]    & 
          0.4629 &    [Ni II]  & 
  **      0.4629 &    [Ni II]  & 
          0.4631 &    [Co III] \\
 ***      0.4641 &    [Fe II]  & 
  **      0.4659 &    [Co II]  & 
          0.4660 &    [Fe III] & 
  **      0.4666 &    [Fe II]  & 
          0.4668 &    [Fe III] \\
   *      0.4685 &    [Co I]   & 
          0.4703 &    [Fe III] & 
  **      0.4712 &    [Ni I]   & 
          0.4715 &    [Co III] & 
   *      0.4719 &    [Co III] \\
 ***      0.4729 &    [Fe II]  & 
          0.4735 &    [Co II]  & 
          0.4735 &    [Fe III] & 
   *      0.4738 &    [Co I]   & 
   *      0.4747 &    [Fe II]  \\
  **      0.4749 &    [Co II]  & 
          0.4756 &    [Fe III] & 
          0.4771 &    [Fe III] & 
  **      0.4773 &    [Fe II]  & 
 ***      0.4776 &    [Fe II]  \\
          0.4779 &    [Fe III] & 
          0.4794 &    [Co II]  & 
  **      0.4800 &    [Fe II]  & 
  **      0.4804 &    [Co II]  & 
   *      0.4815 &    [Ni I]   \\
 ***      0.4816 &    [Fe II]  & 
          0.4818 &    [Co II]  & 
   *      0.4837 &    [Co II]  & 
          0.4841 &    [Co II]  & 
   *      0.4854 &    [Fe II]  \\
          0.4875 &    [Co I]   & 
 ***      0.4876 &    [Fe II]  & 
   *      0.4880 &    [Co II]  & 
 ***      0.4891 &    [Fe II]  & 
   *      0.4899 &    [Co II]  \\
  **      0.4900 &    [Fe II]  & 
 ***      0.4907 &    [Fe II]  & 
   *      0.4910 &    [Co II]  & 
   *      0.4917 &    [Co III] & 
   *      0.4920 &    [Co II]  \\
          0.4932 &    [Fe III] & 
          0.4933 &    [O III]  & 
  **      0.4949 &    [Fe II]  & 
  **      0.4952 &    [Fe II]  & 
 ***      0.4960 &    [O III]  \\
   *      0.4960 &    [O III]  & 
   *      0.4972 &    [Co II]  & 
          0.4972 &    [Co I]   & 
  **      0.4975 &    [Fe II]  & 
          0.4989 &    [Co II]  \\
          0.4989 &    [Co III] & 
  **      0.5007 &    [Fe II]  & 
  **      0.5008 &    [Fe II]  & 
 ***      0.5008 &    [O III]  & 
          0.5013 &    [Fe III] \\
   *      0.5016 &    [Co II]  & 
  **      0.5022 &    [Fe II]  & 
   *      0.5029 &    [Ni I]   & 
   *      0.5029 &    [Fe II]  & 
   *      0.5031 &    [Co II]  \\
   *      0.5037 &    [Fe II]  & 
          0.5045 &    [Co II]  & 
  **      0.5045 &    [Fe II]  & 
   *      0.5050 &    [Fe II]  & 
  **      0.5062 &    [Fe II]  \\
          0.5066 &    [Ni II]  & 
          0.5070 &    [Co I]   & 
  **      0.5074 &    [Fe II]  & 
   *      0.5076 &    [Co I]   & 
          0.5076 &    [Co I]   \\
          0.5080 &    [Co I]   & 
   *      0.5081 &    [Co II]  & 
          0.5086 &    [Fe III] & 
  **      0.5109 &    [Fe II]  & 
 ***      0.5113 &    [Fe II]  \\
   *      0.5115 &    [Co III] & 
          0.5116 &    [Co II]  & 
   *      0.5116 &    [Co II]  & 
   *      0.5125 &    [Co II]  & 
          0.5134 &    [Ni II]  \\
   *      0.5137 &    [Co III] & 
   *      0.5149 &    [Co II]  & 
          0.5149 &    [Co II]  & 
 ***      0.5160 &    [Fe II]  & 
 ***      0.5160 &    [Fe II]  \\
  **      0.5165 &    [Fe II]  & 
   *      0.5167 &    [Co III] & 
   *      0.5174 &    [Fe II]  & 
   *      0.5177 &    [Co II]  & 
  **      0.5183 &    [Fe II]  \\
  **      0.5186 &    [Fe II]  & 
   *      0.5187 &    [Fe II]  & 
          0.5193 &    [Co II]  & 
   *      0.5193 &    [Co III] & 
   *      0.5193 &    [Co III] \\
 ***      0.5221 &    [Fe II]  & 
   *      0.5229 &    [Co II]  & 
          0.5237 &    [Co I]   & 
          0.5237 &    [Co I]   & 
          0.5245 &    [Co I]   \\
          0.5246 &    [Co II]  & 
          0.5249 &    [Co III] & 
          0.5249 &    [Co III] & 
          0.5252 &    [Co I]   & 
          0.5253 &    [Co II]  \\
          0.5253 &    [Co II]  & 
 ***      0.5263 &    [Fe II]  & 
          0.5265 &    [Co I]   & 
 ***      0.5270 &    [Fe II]  & 
  **      0.5270 &    [Co II]  \\
   *      0.5271 &    [Ni II]  & 
          0.5272 &    [Fe III] & 
 ***      0.5275 &    [Fe II]  & 
   *      0.5276 &    [Ni II]  & 
   *      0.5277 &    [Ni II]  \\
          0.5279 &    [Co II]  & 
          0.5283 &    [Ni II]  & 
 ***      0.5298 &    [Fe II]  & 
          0.5299 &    [Co II]  & 
   *      0.5300 &    [Co II]  \\
          0.5330 &    [Co II]  & 
 ***      0.5335 &    [Fe II]  & 
   *      0.5335 &    [Co III] & 
  **      0.5349 &    [Fe II]  & 
 ***      0.5350 &    [Ni I]   \\
          0.5350 &    [Co I]   & 
   *      0.5359 &    [Co II]  & 
          0.5367 &    [Co II]  & 
 ***      0.5378 &    [Fe II]  & 
          0.5388 &    [Co II]  \\
          0.5400 &    [Co I]   & 
          0.5414 &    [Fe III] & 
          0.5416 &    [Co II]  & 
          0.5416 &    [Co II]  & 
          0.5416 &    [Co I]   \\
          0.5433 &    [Ni II]  & 
 ***      0.5435 &    [Fe II]  & 
   *      0.5439 &    [Co II]  & 
          0.5448 &    [Co II]  & 
          0.5448 &    [Co II]  \\
   *      0.5456 &    [Co III] & 
          0.5456 &    [Co III] & 
  **      0.5472 &    [Co II]  & 
  **      0.5479 &    [Fe II]  & 
          0.5515 &    [Co I]   \\
   *      0.5528 &    [Co II]  & 
  **      0.5529 &    [Fe II]  & 
   *      0.5546 &    [Co II]  & 
  **      0.5548 &    [Co II]  & 
  **      0.5553 &    [Fe II]  \\
  **      0.5558 &    [Fe II]  & 
  **      0.5562 &    [Co II]  & 
   *      0.5572 &    [Co II]  & 
   *      0.5574 &    [Co II]  & 
 ***      0.5579 &    [O I]    \\
  **      0.5582 &    [Fe II]  & 
   *      0.5588 &    [Fe II]  & 
   *      0.5589 &    [Fe II]  & 
   *      0.5590 &    [Fe II]  & 
   *      0.5601 &    [Co II]  \\
   *      0.5615 &    [Fe II]  & 
          0.5620 &    [Co II]  & 
          0.5623 &    [Co I]   & 
          0.5625 &    [Co I]   & 
   *      0.5629 &    [Co III] \\
   *      0.5629 &    [Fe II]  & 
  **      0.5641 &    [Fe I]   & 
   *      0.5651 &    [Fe II]  & 
   *      0.5652 &    [Fe II]  & 
   *      0.5659 &    [Co II]  \\
          0.5665 &    [Co II]  & 
          0.5670 &    [Co I]   & 
   *      0.5670 &    [Mn I]   & 
  **      0.5675 &    [Fe II]  & 
   *      0.5676 &    [Co II]  \\
          0.5680 &    [Co II]  & 
  **      0.5683 &    [Co II]  & 
          0.5690 &    [Co I]   & 
   *      0.5692 &    [Mn I]   & 
  **      0.5698 &    [Fe I]   \\
          0.5705 &    [Ni II]  & 
  **      0.5711 &    [Fe I]   & 
          0.5713 &    [Ni II]  & 
   *      0.5720 &    [Fe II]  & 
   *      0.5726 &    [Co II]  \\
   *      0.5727 &    [Fe II]  & 
  **      0.5730 &    [Mn I]   & 
  **      0.5731 &    [Co II]  & 
   *      0.5746 &    [Co II]  & 
  **      0.5749 &    [Fe II]  \\
          0.5758 &    [Co I]   & 
   *      0.5758 &    [Fe II]  & 
   *      0.5769 &    [Fe II]  & 
   *      0.5777 &    [Fe I]   & 
  **      0.5786 &    [Mn I]   \\
   *      0.5801 &    [Fe II]  & 
          0.5803 &    [Co I]   & 
  **      0.5806 &    [Fe I]   & 
   *      0.5816 &    [Co II]  & 
  **      0.5836 &    [Fe I]   \\
  **      0.5837 &    [Fe II]  & 
          0.5841 &    [Co III] & 
   *      0.5846 &    [Fe II]  & 
   *      0.5849 &    [Fe II]  & 
  **      0.5864 &    [Mn I]   \\
   *      0.5869 &    [Fe I]   & 
  **      0.5872 &    [Fe II]  & 
  **      0.5874 &    [Fe I]   & 
          0.5888 &    [Co I]   & 
 ***      0.5890 &    [Co III] \\
          0.5892 &    [Co II]  & 
          0.5894 &    [Co I]   & 
   *      0.5903 &    [Fe II]  & 
  **      0.5908 &    [Co III] & 
   *      0.5915 &    [Fe II]  \\
\hline
\end{tabular}
\end{center}
\label{tab:table2b}
\end{table*}

\begin{table*}
\begin{center}
    \caption{Same as Table \ref{table.prediction} but for synthetic spectra in the optical (Part 3). Note that the [CaII] doublet at $0.7293/
    0.7326 \mu$m is dominating beyond scale. Without Ca, there would still be  significant features comparable to those at shorter wavelengths as is obvious from the red 'component' of model 03 (see inset of Fig. \ref{IR_spectrum}).}
    \begin{tabular}{|rr|rr|rr|rr|rr|}
    \hline    
    S~~~$\lambda $[\microns] &Ion  &   
    S~~~$\lambda $[\microns] &Ion  &   
    S~~~$\lambda $[\microns] &Ion  &   
    S~~~$\lambda $[\microns] &Ion  &   
    S~~~$\lambda $[\microns] &Ion  \\
\hline
          0.5925 &    [Co II]  & 
  **      0.5936 &    [Fe I]   & 
  **      0.5939 &    [Fe I]   & 
          0.5943 &    [Co III] & 
          0.5951 &    [Co II]  \\
   *      0.5956 &    [Co II]  & 
   *      0.5960 &    [Co I]   & 
   *      0.5961 &    [Co II]  & 
  **      0.5971 &    [Fe I]   & 
   *      0.5984 &    [Fe II]  \\
          0.5989 &    [Co II]  & 
          0.5989 &    [Co II]  & 
   *      0.6002 &    [Fe I]   & 
   *      0.6009 &    [Ni II]  & 
          0.6016 &    [Co II]  \\
          0.6020 &    [Co II]  & 
          0.6022 &    [Co II]  & 
          0.6031 &    [Co I]   & 
  **      0.6046 &    [Fe II]  & 
          0.6051 &    [Co I]   \\
          0.6058 &    [Co II]  & 
          0.6059 &    [Co II]  & 
   *      0.6072 &    [Co I]   & 
          0.6075 &    [Co II]  & 
   *      0.6091 &    [Co I]   \\
   *      0.6097 &    [Fe II]  & 
          0.6113 &    [Co II]  & 
          0.6113 &    [Co I]   & 
  **      0.6129 &    [Co III] & 
   *      0.6149 &    [Co II]  \\
   *      0.6154 &    [Co II]  & 
   *      0.6187 &    [Co II]  & 
          0.6188 &    [Co II]  & 
   *      0.6190 &    [Fe II]  & 
  **      0.6190 &    [Fe II]  \\
  **      0.6197 &    [Co III] & 
          0.6210 &    [Co III] & 
   *      0.6247 &    [Co II]  & 
          0.6253 &    [Co II]  & 
          0.6258 &    [Co II]  \\
   *      0.6263 &    [Fe II]  & 
          0.6265 &    [Co I]   & 
   *      0.6265 &    [Co I]   & 
   *      0.6276 &    [Co II]  & 
   *      0.6282 &    [Fe II]  \\
   *      0.6289 &    [Co II]  & 
          0.6295 &    [Co II]  & 
 ***      0.6302 &    [O I]    & 
   *      0.6302 &    [O I]    & 
          0.6303 &    [Co II]  \\
   *      0.6316 &    [Co II]  & 
   *      0.6318 &    [Co I]   & 
          0.6343 &    [Co I]   & 
          0.6352 &    [Co I]   & 
          0.6352 &    [Co I]   \\
   *      0.6355 &    [Fe II]  & 
          0.6356 &    [Co II]  & 
          0.6356 &    [Co II]  & 
          0.6365 &    [Co I]   & 
 ***      0.6366 &    [O I]    \\
   *      0.6366 &    [O I]    & 
   *      0.6367 &    [Ni II]  & 
   *      0.6388 &    [Co I]   & 
   *      0.6389 &    [Co I]   & 
          0.6393 &    [O I]    \\
   *      0.6398 &    [Fe II]  & 
          0.6400 &    [Co II]  & 
  **      0.6406 &    [Ni I]   & 
   *      0.6406 &    [Fe II]  & 
          0.6426 &    [Co II]  \\
  **      0.6439 &    [Ni I]   & 
   *      0.6443 &    [Ni II]  & 
   *      0.6469 &    [Ni II]  & 
          0.6471 &    [Co II]  & 
   *      0.6476 &    [Fe II]  \\
   *      0.6476 &    [Fe II]  & 
   *      0.6484 &    [Fe II]  & 
   *      0.6487 &    [Fe II]  & 
  **      0.6491 &    [Ni I]   & 
          0.6506 &    [Co I]   \\
          0.6506 &    [Co I]   & 
   *      0.6513 &    [Fe II]  & 
  **      0.6521 &    [Co II]  & 
   *      0.6529 &    [Si I]   & 
   *      0.6547 &    [Fe II]  \\
   *      0.6568 &    [Fe II]  & 
  **      0.6578 &    [Co III] & 
   *      0.6586 &    [Co I]   & 
   *      0.6586 &    [Fe II]  & 
          0.6591 &    [Si I]   \\
  **      0.6606 &    [Ni I]   & 
   *      0.6634 &    [Co II]  & 
          0.6639 &    [Co I]   & 
   *      0.6639 &    [Co I]   & 
          0.6652 &    [Co I]   \\
   *      0.6669 &    [Ni II]  & 
   *      0.6674 &    [Fe II]  & 
          0.6680 &    [Co I]   & 
          0.6680 &    [Co I]   & 
   *      0.6685 &    [Co I]   \\
   *      0.6691 &    [Fe II]  & 
          0.6702 &    [Ni II]  & 
   *      0.6702 &    [Fe II]  & 
          0.6707 &    [Co II]  & 
   *      0.6710 &    [Mn II]   \\
   *      0.6718 &    [S II]   & 
  **      0.6732 &    [Fe II]  & 
   *      0.6732 &    [Ni I]   & 
   *      0.6733 &    [S II]   & 
   *      0.6733 &    [Co II]  \\
          0.6747 &    [Co II]  & 
   *      0.6748 &    [Fe II]  & 
   *      0.6749 &    [Fe II]  & 
   *      0.6762 &    [Co I]   & 
          0.6765 &    [Mn II]   \\
   *      0.6789 &    [Ni I]   & 
          0.6792 &    [Co I]   & 
   *      0.6793 &    [Ni II]  & 
   *      0.6796 &    [Ni II]  & 
          0.6799 &    [Co II]  \\
          0.6806 &    [Co I]   & 
  **      0.6811 &    [Fe II]  & 
   *      0.6815 &    [Ni II]  & 
          0.6821 &    [Co I]   & 
          0.6830 &    [Co II]  \\
   *      0.6831 &    [Fe II]  &
   *      0.6850 &    [Ni II]  &
   *       0.6852 &    [Mn II]  & 
   *      0.6851 &    [Co III] & 
   **      0.6855 &    [Co III] \\
  **      0.6874 &    [Fe II]  & 
  **      0.6876 &    [Fe II]  & 
          0.6878 &    [Co II]  & 
   *      0.6888 &    [Co II]  & 
  **      0.6898 &    [Fe II]  \\
          0.6907 &    [Co I]   & 
   *      0.6913 &    [Ni II]  & 
          0.6925 &    [Fe II]  & 
   *      0.6926 &    [Co II]  & 
  **      0.6934 &    [Co II]  \\
   *      0.6936 &    [Fe II]  & 
  **      0.6944 &    [Ni I]   & 
  **      0.6947 &    [Fe II]  & 
   *      0.6951 &    [Co I]   & 
   *      0.6958 &    [Ni II]  \\
          0.6958 &    [Ni II]  & 
   *      0.6963 &    [Co III] & 
   *      0.6964 &    [Co III] & 
  **      0.6968 &    [Fe II]  & 
   *      0.6974 &    [Co I]   \\
   *      0.6981 &    [Co II]  &
  **      0.7004 &    [Ni I]    &
   *      0.7012 &    [Co II]  & 
   *      0.7013 &    [Fe II]  & 
          0.7014 &    [Co III] \\
          0.7019 &    [Co I]   & 
          0.7030 &    [Co II]  & 
   *      0.7031 &    [Co II]  & 
          0.7037 &    [Co II]  &
   *      0.7050 &    [Fe II]  \\ 
   *      0.7052 &    [Co II]  & 
   *      0.7053 &    [Co II]  & 
   *      0.7056 &    [Ni II]  & 
   *      0.7080 &    [Ni II]  &
          0.7100 &    [Co I]   \\ 
   *      0.7102 &    [Co II]  & 
          0.7105 &    [Ni II]  & 
  **      0.7132 &    [Ni I]   & 
   *      0.7133 &    [Fe II]  &
   *      0.7135 &    [Co I]   \\
   *      0.7142 &    [Co II]  & 
          0.7145 &    [Co II]  & 
          0.7147 &    [Co II]  & 
   *      0.7155 &    [Co III] &
 ***      0.7157 &    [Fe II]  \\ 
   *      0.7161 &    [Co I]   &
          0.7162 &    [Co III] &
   *      0.7171 &    [Co III] & 
          0.7173 &    [Co II]  & 
 ***      0.7174 &    [Fe II]  \\ 
  **      0.7196 &    [Ni I]   & 
          0.7204 &    [Co III] & 
   *      0.7221 &    [Ni I]   & 
   *      0.7249 &    [Co I]   & 
   *      0.7258 &    [Ni II]  \\
   *      0.7258 &    [Co II]  &
          0.7258 &    [Co I]   & 
   *      0.7272 &    [Co III] & 
   *      0.7277 &    [Co II]  & 
 ****      0.7293 &    [Ca II]  \\ 
          0.7300 &    [Co I]   &
          0.7300 &    [Co I]   & 
   *      0.7310 &    [Ni II]  & 
   *      0.7321 &    [O II]   & 
  **      0.7322 &    [O II]   \\ 
 ****      0.7326 &    [Ca II]  &
   *      0.7332 &    [O II]   & 
   *      0.7332 &    [O II]   & 
   *      0.7332 &    [Fe II]  & 
   *      0.7333 &    [O II]   \\ 
   *      0.7333 &    [O II]   &
          0.7360 &    [Co III] & 
          0.7365 &    [Co II]  & 
   *      0.7372 &    [Co II]  & 
   *      0.7373 &    [Fe II]  \\
          0.7373 &    [Co II]  & 
   *      0.7374 &    [Co III] & 
   *      0.7375 &    [Co II]  & 
   *      0.7380 &    [Ni II]  &
 ***      0.7390 &    [Fe II]  \\ 
          0.7394 &    [Co II]  & 
   *      0.7396 &    [Ni I]   & 
   *      0.7398 &    [Ni I]   & 
   *      0.7410 &    [Co II]  & 
          0.7414 &    [Ni II]  \\ 
   *      0.7424 &    [Co II]  & 
          0.7430 &    [Co I]   & 
   *      0.7434 &    [Fe II]  &
          0.7447 &    [Co II]  & 
 ***      0.7455 &    [Fe II]  \\ 
\hline
\end{tabular}
\end{center}
\label{tab:table2c}
\end{table*}

\begin{table*}
\begin{center}
    \caption{Same as Table \ref{table.prediction} but for synthetic spectra in the optical (Part 4).}
    \begin{tabular}{|rr|rr|rr|rr|rr|}
    \hline
        S~~~$\lambda $[\microns] &Ion  &   
    S~~~$\lambda $[\microns] &Ion  &   
    S~~~$\lambda $[\microns] &Ion  &   
    S~~~$\lambda $[\microns] &Ion  &   
    S~~~$\lambda $[\microns] &Ion  \\
\hline
   *      0.7466 &    [Ni I]   & 
   *      0.7469 &    [Co II]  & 
          0.7477 &    [Co I]   & 
   *      0.7482 &    [Co II]  & 
   *      0.7499 &    [Co III] \\
  **      0.7509 &    [Ni I]   & 
          0.7511 &    [Co I]   & 
   *      0.7521 &    [Co I]   & 
          0.7522 &    [Co II]  &
          0.7526 &    [Co II]  \\ 
  **      0.7541 &    [Co II]  &
   *      0.7556 &    [Co II]  & 
   *      0.7564 &    [Co I]   &
   *      0.7570 &    [Co II]  &
          0.7584 &    [Co II]  \\ 
   *      0.7614 &    [Co II]  & 
   *      0.7615 &    [Ni II]  &
   *      0.7615 &    [Fe II]  &
          0.7639 &    [Co I]   &
          0.7639 &    [Co I]   \\ 
  **      0.7640 &    [Fe II]  & 
   *      0.7640 &    [Co II]  &
   *      0.7645 &    [Co II]  &
   *      0.7658 &    [Co I]   & 
   *      0.7667 &    [Fe II]  \\
  **      0.7689 &    [Fe II]  &
          0.7697 &    [Ni II]  &
          0.7701 &    [Co I]   & 
   *      0.7704 &    [Co I]   &
          0.7706 &    [Co II]  \\ 
   *      0.7722 &    [Fe II]  &
   *      0.7725 &    [Co II]  &
 ***      0.7727 &    [S I]    & 
          0.7731 &    [Co II]  &
   *      0.7735 &    [Fe II]  \\ 
   *      0.7737 &    [Fe II]  &
          0.7757 &    [Co II]  &
  **      0.7767 &    [Fe II]  & 
   *      0.7796 &    [Co II]  &
          0.7822 &    [Co III] \\
          0.7832 &    [Co II]  &
   *      0.7837 &    [Co II]  &
   *      0.7840 &    [Co I]   & 
   *      0.7876 &    [Fe II]  &
  **      0.7910 &    [Ni I]   \\
   *      0.7914 &    [Co I]   & 
   *      0.7918 &    [Co I]   & 
   *      0.7926 &    [Co II]  & 
          0.7932 &    [Ni I]   &
   *      0.7936 &    [Co II]  \\
          0.7936 &    [Co II]  & 
   *      0.7943 &    [Co I]   & 
          0.7972 &    [Co I]   & 
   *      0.7977 &    [Fe II]  &
          0.7992 &    [Ni I]   \\
   *      0.8030 &    [Co II]  & 
   *      0.8036 &    [Ni II]  & 
          0.8036 &    [Ni II]  & 
   *      0.8047 &    [Fe II]  &
   *      0.8049 &    [Co II]  \\
          0.8072 &    [Co I]   & 
   *      0.8080 &    [Fe II]  & 
   *      0.8112 &    [Co I]   &
   *      0.8114 &    [Ni I]   & 
  **      0.8123 &    [Co II]  \\ 
          0.8179 &    [Co I]   & 
          0.8180 &    [Co II]  &
   *      0.8197 &    [Ni I]   &
   *      0.8199 &    [Co II]  & 
   *      0.8201 &    [Fe II]  \\ 
          0.8204 &    [Ni I]   & 
   *      0.8262 &    [Fe II]  &
          0.8273 &    [Co I]   &
   *      0.8303 &    [Ni II]  & 
          0.8306 &    [Co I]   \\ 
  **      0.8308 &    [Fe II]  & 
   *      0.8336 &    [Co II]  &
          0.8343 &    [Co I]   &
  **      0.8389 &    [Fe II]  & 
   *      0.8413 &    [Co II]  \\ 
   *      0.8414 &    [Fe II]  &
   *      0.8466 &    [Co II]  &
   *      0.8469 &    [Ni I]   &
   *      0.8469 &    [Co II]  & 
   *      0.8482 &    [Fe II]  \\ 
   *      0.8489 &    [Co I]   &
   *      0.8492 &    [Fe II]  &
          0.8500 &    [Co II]  & 
   *      0.8543 &    [Co II]  & 
          0.8545 &    [Co I]   \\
   *      0.8546 &    [Co I]   &
  **      0.8574 &    [Co II]  &
   *      0.8577 &    [Fe II]  & 
   *      0.8578 &    [Fe II]  & 
   *      0.8583 &    [Co II]  \\
   *      0.8597 &    [Co I]   &
   *      0.8602 &    [Co II]  &
  **      0.8603 &    [Fe II]  & 
 ***      0.8619 &    [Fe II]  & 
   *      0.8626 &    [Co II]  \\
   *      0.8691 &    [Co I]   &
          0.8694 &    [Co II]  & 
   *      0.8706 &    [Ni II]  & 
   *      0.8709 &    [Co II]  &
   *      0.8711 &    [Fe II]  \\
  **      0.8718 &    [Fe II]  &
 ***      0.8730 &    [C I]    & 
   *      0.8737 &    [Fe II]  &
   *      0.8741 &    [Fe II]  &
          0.8776 &    [Co I]   \\ 
   *      0.8827 &    [Fe II]  &
  **      0.8835 &    [Ni I]   &
  **      0.8846 &    [Ni I]   & 
   *      0.8855 &    [Co II]  &
   *      0.8864 &    [Fe II]  \\
   *      0.8888 &    [Fe II]  &
  **      0.8894 &    [Fe II]  &
          0.8899 &    [Ni II]  & 
          0.8899 &    [Co II]  &
   *      0.8934 &    [Fe II]  \\
   *      0.8968 &    [Co II]  &
   *      0.8993 &    [Co I]   &
   *      0.9015 &    [Co II]  & 
   *      0.9022 &    [Co I]   &
  **      0.9036 &    [Fe II]  \\
  **      0.9054 &    [Fe II]  &
   *      0.9057 &    [Co I]   & 
   *      0.9085 &    [Co II]  &
   *      0.9086 &    [Fe II]  &
   *      0.9135 &    [Co I]   \\
   *      0.9136 &    [Fe II]  &
          0.9149 &    [Co II]  & 
   *      0.9205 &    [Fe II]  & 
   *      0.9205 &    [Co II]  &
  **      0.9229 &    [Fe II]  \\
  **      0.9234 &    [Fe II]  & 
   *      0.9247 &    [Co II]  & 
  **      0.9270 &    [Fe II]  & 
  **      0.9338 &    [Co II]  &
 ***      0.9345 &    [Co II]  \\
          0.9347 &    [Co I]   &
   *      0.9354 &    [Fe II]  & 
   *      0.9372 &    [Co II]  &
          0.9377 &    [Ni II]  & 
   *      0.9384 &    [Fe II]  \\
   *      0.9387 &    [Fe II]  &
   *      0.9414 &    [Fe I]   & 
   *      0.9439 &    [Fe II]  &
          0.9447 &    [Fe III] & 
   *      0.9468 &    [Fe II]  \\
   *      0.9472 &    [Fe II]  &
  **      0.9474 &    [Fe II]  & 
   *      0.9493 &    [Fe II]  &
   *      0.9517 &    [Fe II]  & 
          0.9529 &    [Co I]   \\
   *      0.9541 &    [Co II]  &
   *      0.9555 &    [Fe II]  &
   *      0.9593 &    [Fe II]  & 
  **      0.9642 &    [Co II]  & 
   *      0.9662 &    [Fe I]   \\
   *      0.9672 &    [Fe II]  & 
   *      0.9685 &    [Fe II]  &
   *      0.9696 &    [Co I]   & 
          0.9704 &    [Fe III] & 
   *      0.9714 &    [Fe II]  \\
          0.9760 &    [Ni II]  &
   *      0.9805 &    [Fe I]   & 
  **      0.9827 &      [C I]  &
   *      0.9829 &    [Fe I]   &
  **      0.9853 &    [C I]    \\ 
\hline
\end{tabular}
\end{center}
\label{tab:table2d}
\end{table*}

\end{appendix}

\end{document}